\documentclass[11pt, conference]{IEEEtran}
\IEEEoverridecommandlockouts
\renewcommand\IEEEkeywordsname{Keywords}

\usepackage{cite}
\usepackage{amsmath,amssymb,amsfonts}
\usepackage{algorithmic}
\usepackage{graphicx}
\usepackage{textcomp}
\usepackage{xcolor}
\usepackage{afterpage}

\usepackage{comment}
\usepackage{graphicx} %to include picture
\usepackage[colorinlistoftodos]{todonotes}
\usepackage[utf8]{inputenc}
\usepackage{multirow}
\usepackage{multicol}
\usepackage{slashbox}
\usepackage{caption}
\usepackage{subcaption}
\newcommand{\RomanNumeralCaps}[1]
    {\MakeUppercase{\romannumeral #1}}
\usepackage{authblk}
\usepackage{float} 
\usepackage{balance}
\usepackage{fancyhdr}

\def\BibTeX{{\rm B\kern-.05em{\sc i\kern-.025em b}\kern-.08em
    T\kern-.1667em\lower.7ex\hbox{E}\kern-.125emX}}

\usepackage[english]{babel}
\usepackage[utf8]{inputenc}
\usepackage{fancyhdr}

\begin{document}

\title{Quantifiable Assurance: From IPs to Platforms\\

}

\author{Bulbul Ahmed}
\author{Md Kawser Bepary}
\author{Nitin Pundir}
\author{Mike Borza}
\author{Oleg Raikhman}
\author{Amit Garg}
\author{\\Dale Donchin}
\author{Adam Cron}
\author{Mohamed A Abdel-moneum}

%\author[$\dagger$]{Nick Hooten}
\author{Farimah Farahmandi}
\author{Fahim Rahman}
\author{\\Mark Tehranipoor}

\affil{Email: \{ahmed.b, mdkawser.bepary, nitin.pundir\}@ufl.edu, Department of ECE, University of Florida \\
Email: \{mborza, oraikhm, amitgarg, dale.donchin, acron\}@synopsys.com, Synopsys Inc.\\
Email: mohamed.a.abdel-moneum@intel.com, Intel Corporation.\\
Email: \{farimah, fahimrahman, tehranipoor\}@ece.ufl.edu, Department of ECE, University of Florida
}

\maketitle
\thispagestyle{fancy}

%\balance
\begin{abstract}

Hardware vulnerabilities are generally considered more difficult to fix than software ones because they are persistent after fabrication. Thus, it is crucial to assess the security and fix the vulnerabilities at earlier design phases, such as Register Transfer Level (RTL) and gate level. The focus of the existing security assessment techniques is mainly twofold. First, they check the security of Intellectual Property (IP) blocks separately. Second, they aim to assess the security against individual threats considering the threats are orthogonal. We argue that IP-level security assessment is not sufficient. Eventually, the IPs are placed in a platform, such as a system-on-chip (SoC), where each IP is surrounded by other IPs connected through glue logic and shared/private buses. Hence, we must develop a methodology to assess the platform-level security by considering both the IP-level security and the impact of the additional parameters introduced during platform integration. Another important factor to consider is that the threats are not always orthogonal. Improving security against one threat may affect the security against other threats. Hence, to build a secure platform, we must first answer the following questions: What additional parameters are introduced during the platform integration? How do we define and characterize the impact of these parameters on security? How do the mitigation techniques of one threat impact others? This paper aims to answer these important questions and proposes techniques for quantifiable assurance by quantitatively estimating and measuring the security of a platform at the pre-silicon stages. We also touch upon the term security optimization and present the challenges for future research directions.


\end{abstract}

\section{Introduction}
System-on-Chips (SoCs) have been pervasive in current and future electronic products. %(Note that we will interchangeably use SoC and platform in this paper). 
Almost every computing system (e.g., mobile phones, payment gateways, IoT devices, medical equipment, automotive systems, avionic devices, etc.) is built around the SoC. A modern SoC contains a wide range of sensitive information, commonly referred to as security assets (e.g., keys, biometrics, personal info, etc.). The assets are necessary to build security mechanisms and need to be protected from a diverse set of attack models, such as IP piracy \cite{yasin2016sarlock}, power side channel analysis \cite{kocher1999differential}, , fault-injection, malicious hardware attack \cite{tehranipoor2010survey}, supply chain attack \cite{guin2014low} etc. 

Often, hardware designs cannot be patched to fix security vulnerabilities after the design is fabricated and deployed in the field. Any changes to the hardware design after fabrication would require redoing the entire design, which costs money and time in addition to losing market share. Hence, it is crucial to fix the vulnerabilities at the earlier design phases, such as RTL or gate-level. Additionally, a quantifiable assurance, such as quantitative estimation and measurement of the security, is much needed to evaluate the security of the whole design.

\begin{figure}[t]
 \vspace{0.5em}
    \centering
    \includegraphics[width=0.6\columnwidth]{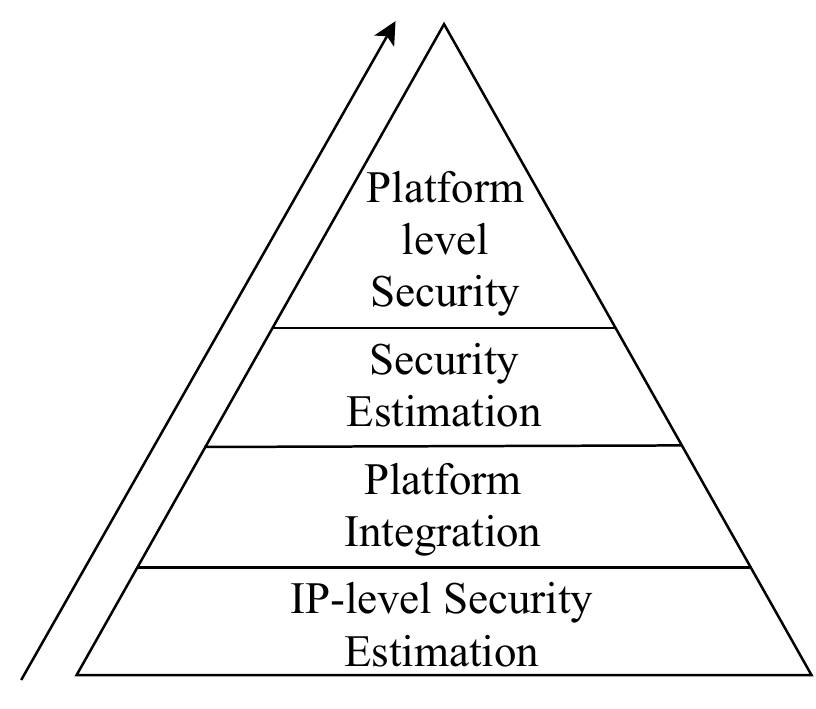}
 %   \vspace{-1em}
  %\vspace{-1.5em}
    \caption{The approach to platform-level security estimation.}
    \label{fig:pyramidApproach}
   \vspace{-1.5em}
\end{figure}

While much research has been done in the testing domain \cite{ahmed2006novel, ahmed2005speed, tehranipoor2011test, yilmaz2010test, lee2008layout, ahmed2007supply, tehranipour2003testing, ahmed2004low, ma2009layout}, security verification is still on the rise, and the modern Electronic Design Automation (EDA) tools still lack the notion of automatic security evaluation and optimization. In most cases, security is considered an afterthought.  To address this, researchers have developed different detection \cite{salmani2010layout, lamech2011experimental,ahi2015terahertz, villasenor2013chop} and defense \cite{karimian2016highly, rahman2014csst, rahman2019dynamically, nahiyan2018security, zhou2021data,9586106,sami_poca_2021} mechanisms to build secure designs. At the same time, several methodologies have been developed by the hardware security community to assess the defense mechanisms quantitatively. However, all these approaches can be only applied to intellectual property (IP) blocks, which are the primary building blocks of the platform. Note that the terms "System-on-Chip (SoC) and "Platform" will be used interchangeably in the rest of the paper. The security of an IP does not necessarily remain the same after it is integrated into the platform. During the integration, additional parameters are introduced that directly impact the security of an IP at the platform level. These parameters either degrade or upgrade the security of the IP after being placed in the platform. It leads to the development of new platform-level security estimation and measurement methodologies. While the security estimation methods should accurately estimate the silicon-level security of the platform at the earlier design stages, the measurement approach should accurately measure the platform-level security by exhaustive simulation or emulation.Another important aspect is that threats are usually considered individually during mitigation. Mitigating vulnerabilities for one threat can potentially make the design more vulnerable to another threat. Thus, security optimization is needed to obtain the best possible collective security considering the diverse threat model. A high-level overview of the platform-level security estimation approach is shown in Figure \ref{fig:pyramidApproach}. It follows the bottom-up approach starting with the IP-level security estimation following the platform integration and the platform-level security estimation.

To the best of our knowledge, this paper is the first to propose a comprehensive approach to estimate and measure the platform-level security at the earlier phases of a design. We first discuss the IP-level security metrics and different IP-level parameters that contribute to the security at the IP-level for five different threats: IP Piracy, Power Side-Channel Analysis, Fault-Injection, Malicious Hardware, and Supply Chain. Then, we discuss the transition from IP to platform and show different parameters introduced during this transition. Then, we describe our proposed approach for the platform-level security estimation and measurement. We also apply our proposed techniques on two case studies for the estimation and measurement of resiliency against IP Piracy and the Power Side-Channel (PSC) analysis attacks at the platform level.

This paper is organized by starting with a motivating example in Section \RomanNumeralCaps{2}. Then, in Section \RomanNumeralCaps{3}, we discuss the threat models. In Section \RomanNumeralCaps{4}, we present the background of our work, including the discussion on IP-level security metrics and design parameters contributing to the platform-level security metric. Section \RomanNumeralCaps{5} describes the transition from IP to SoC and the additional parameters introduced during this transition. The proposed approach for security measurement and estimation is presented in Section \RomanNumeralCaps{7}. We discuss a relatively new concept called Security Optimization in Section \RomanNumeralCaps{8} followed by the challenges in Section \RomanNumeralCaps{9}. Finally, the paper is concluded in Section \RomanNumeralCaps{10}.
%\todo[inline]{Add contribution}
%\todo[inline]{talk about the organization of the paper......}

\begin{figure}[]
 \vspace{0.5em}
    \centering
    \includegraphics[width=0.6\columnwidth]{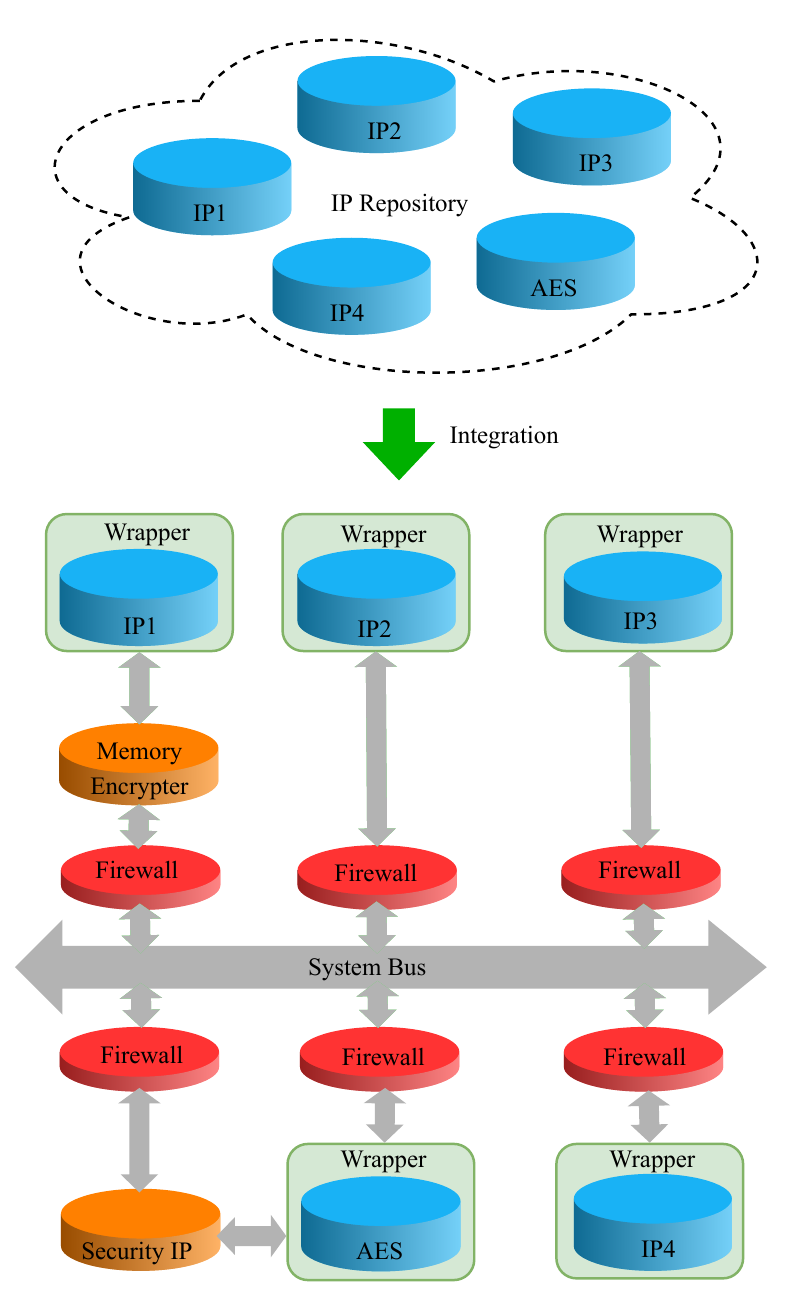}
 %   \vspace{-1em}
  %\vspace{-1.5em}
    \caption{The platform integration flow, the transition from IPs to the platform.}
    \label{fig:ipToSoC}
   \vspace{-1.5em}
\end{figure}

\section{Motivating Example}
This section provides a motivating example and explains why there is a need to develop the SoC-level security estimation and measurement techniques. For this example, we consider the power side-channel (PSC) analysis as the threat model of the interest. Figure \ref{fig:ipToSoC} shows the transition from IP to the platform. Let's have a look at a cryptographic core, IP\textsubscript{2}, shown in Figure \ref{fig:ipToSoC}, assuming that IP\textsubscript{2}'s security against the PSC vulnerability has already been evaluated as a standalone module. Now, we want to check if the IP\textsubscript{2}'s robustness against the PSC  attacks remains the same after the IP is integrated into a system. %the platform is fabricated into silicon?
If not, does it decrease or increase, and by how much? To understand how SoC parameters affect the PSC robustness on IP\textsubscript{2}, let's consider PDN. The first step to perform power side-channel leakage analysis on the crypto core, IP\textsubscript{2}, is to collect the power traces of the IP. 
However, depending on how PDN is implemented at the SoC level, different power rails are shared among different IPs. Therefore, it is impossible to collect power trace of only from IP\textsubscript{2} at the platform level. On the other hand, one assumption while assessing the IP-level PSC robustness is that the attacker has the privilege to measure the power trace of the IP itself. This assumption, however, does not hold true at the platform level since the attacker can only collect the power trace of the entire platform, which is the accumulated power trace of all active IPs along with IP\textsubscript{2}. This might make the collected power traces noisier than the power traces collected from standalone IP\textsubscript{2}, potentially making the attackers' job more challenging. In other words, the security of IP\textsubscript{2} against PSC attacks might be stronger when it is integrated into the SoC unlike what it is measured at the IP-level. 

Thus, the platform-level security of an IP is subject to change depending on different parameters introduced during the transition from IP to platform. A list of such parameters affecting the platform-level security for five different threats is included in Table \ref{tab:ipAndSoC_LevelParam}. An accurate estimation greatly affects the design choice. For example, in the case of PSC assessment, if the estimated SoC-level security of IP2 is high enough, we may not shield the IP through additional countermeasures, which significantly saves design effort, area, power, and cost. This example implies that the transition from IP to platform affects the security at the platform level by introducing different parameters, leading to the need for the platform-level security estimation and measurement methodology.

\section{Threat model}\label{sec:threatModel}
In this section, we  explain the threat models for the five different attacks. We briefly describe how these attacks are performed at the IP level. This section lays a solid foundation for presenting security estimation and measurement methods at the IP and platform levels.

\textbf{IP Piracy:} This attack refers to making an illegal copy or cloning of the IP. Usually, an attacker makes little or no modification to the original IP and sells it to a chip designer claiming its ownership. Logic locking has been a popular technique to protect IP piracy. In this approach, the design is locked by inserting a set of key gates at the gate-level netlist. Only with the correct value of the key inputs, the IP is expected to function correctly. For all other wrong key values, a corrupted output will be produced so that the attacker cannot retrieve the functionality of the IP \cite{lee2005securing, lee2006low, lee2007securing}. In recent years, Boolean Satisfiability Checking-based attack (SAT attack) has been a very efficient technique to retrieve the correct key from a locked gate-level netlist. It has been shown that the SAT-based attack can retrieve the correct key in a few hours for a considerably large keyspace \cite{subramanyan2015evaluating}. This attack utilizes the modern SAT solver and iteratively solves a SAT formula, Conjunctive Normal Format (CNF) of the design,  and prunes out the wrong keys till the correct key is retrieved. The attack model is summarized as follows:
\begin{itemize}
    \item The attacker has access to the locked IP's gate-level netlist.
    \item The attacker also has a functional copy of IC, called \textit{Oracle}, in which she can apply inputs and observe the correct outputs.
    \item The attacker has access to the scan chain of the IC.
\end{itemize}
For sequential designs, it is important to have access to scan chain to be able to perform the SAT attack. Access to the scan chain and the ability to perform the scan shift operation provides full controllability of the circuit’s internal states and reduces the complexity of SAT attacks, just like applying it on a combinational design \cite{el2017reverse}. The SAT solving tools produce $distinguishing$ $input$ $patterns$ $(DIPs)$ \cite{subramanyan2015evaluating} and prune out the wrong keyspace, resulting in finding the correct key at the end. DIP is an input pattern, $x_d$, for which there are at least two different key values, $k_1$, and $k_2$, that generate two different outputs, $o_1$, and $o_2$.
A SAT-solving tool first finds out the DIP, $x_d$, which is then applied to the functional IC to obtain the correct output $o_d$. This correct input-output pair, $(x_d, o_d)$, is then compared to the input-output pair of the locked design. If they do not match, then the key-values used to find the DIP are pruned out. Note that all the wrong keys cannot be pruned out with a single DIP. This process continues until no further DIP is found. Consequently, all the wrong key values will be pruned out, leaving only the correct key.

\textbf{Power Side-Channel (PSC) Leakage: }The threat model for the power side-channel attack assumes that there are distinct assets (keys) in the critical IPs that are subject to the risk of being revealed by the power consumption analysis. Power side-channel attacks utilize the traces of power consumption of an IP to extract the secret key from a cryptographic implementation such as advanced encryption standard (AES) \cite{daemen2002design}. 
In the case of AES algorithm, the secret key is used to generate a number of round keys. The plaintext goes through a series of round operations, and each round consists of a byte substitution, row shifting, mix column and add round key operations. The power consumption of a device is determined by the transistor/switching activity of the chip. However, it turns out that the power consumed at different phases of the cryptographic operation is related to the secret key's value. For example, the s-box computation in the first round allows power traces to be statistically distinguished based on the value of the least significant bit of the s-box output \cite{kocher2011introduction}. 

In the context of platform-level security measurement and estimation, we consider cryptographic IPs where the attacker's goal is to leak the key through differential power analysis (DPA), or correlation power analysis (CPA) attacks \cite{kocher1999differential,brier2004correlation}. The attacker is assumed to have complete access to the power consumption measurement of the target chip.
An adversary begins by collecting power traces during cryptographic operations on the chip for a large set of known plaintexts in order to leak the secret key. The adversary then employs a divide-and-conquer strategy to piece together the secret key byte by byte. The adversary takes into account all 256 hypothetical values for each byte and divides the traces into two sets based on the value of the first s-box output's least significant bit (0 or 1). Then two sets are compared to see if there is a statistical difference; the correct hypothetical key causes the two sets to differ significantly \cite{kocher2011introduction}.
 %The estimation tool requires information to distinguish between cryptographic and non-cryptographic IPs, as well as asset identification, during the estimation process.

\begin{figure}[]
    \centering
    \includegraphics[width=0.9\columnwidth]{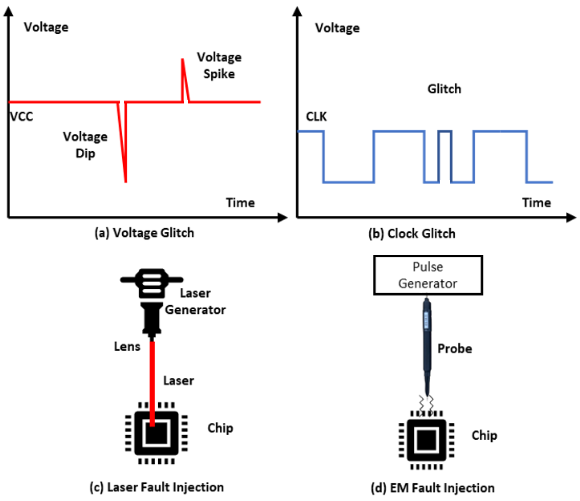}
    \setlength{\belowcaptionskip}{0pt}
    \caption{Different fault-injection techniques.}
    \label{fig:fault_injection_overview}
    \vspace{-1.5em}
\end{figure}

%\color{blue}
\textbf{Fault Injection:} In the hardware domain, fault injection attacks are among the most common non/semi-invasive side-channel attacks \cite{hsueh1997fault} \cite{ziade2004survey}. An attacker creates transient or permanent faults during the device's normal operation and observes the erroneous outputs. It allows an attacker to drastically reduce the number of experiments needed to guess the secret, also known as Differential Fault Analysis (DFA) \cite{dusart2003differential}.

Prominent fault injection techniques include voltage and clock glitching, EM/radiation injection, and laser/optical injection \cite{ziade2004survey}. Voltage glitching involves a momentary power drop or spike in the supply voltage during the operation, as shown in Figure \ref{fig:fault_injection_overview}(a). These voltage spikes or drops affect the critical paths thus causing latching failures of the signals \cite{nahiyan2016avfsm}. A voltage glitch can be caused by either disturbing the main power supply causing global effect, or running a power hungry circuit like ring oscillators (ROs) to cause a localized voltage drop. In recent years, it has been demonstrated that remote fault injection attacks are also possible by disturbing the supply voltage remotely by using either software-based registers to control the supply voltage or using ROs to cause voltage drop in remote FPGAs \cite{murdock2020plundervolt, krautter2018fpgahammer}. Clock glitching involves adding glitches in the clock supply or disturbing normal behavior, as shown in Figure \ref{fig:fault_injection_overview}(b). These clock glitches can cause setup and hold time violations \cite{agoyan2010clocks}. Since voltage starvation can also impact the clock and circuit timing, voltage and clock glitching attacks are combined to cause timing faults.

Optical/laser fault injection attacks use different wavelength laser/optics to inject transient faults from either front or backside of the chip, as shown in Figure \ref{fig:fault_injection_overview}(c). In order to attack from the front-side higher wavelength laser (1300$\mu m$) is used, whereas from the backside near-infrared laser (1064$\mu m$) is used due to its lower absorption coefficient. The injected laser generates electron-hole pairs in the active region, which drift apart under the electric field's influence, resulting in transitory currents. These transient currents cause the transistors to conduct and thus cause the charging/discharging of the capacitive loads. Laser/optical fault injection generally requires de-packaging of the chip to expose the die and thus qualifies as a semi-invasive attack. However, with the laser/optics, an attacker can target the precise locations and time of the laser injection, thus making it a powerful attack \cite{van2011practical}. Similarly, EM fault injection attacks uses electric or magnetic field flux to influence normal functioning of the device, as shown in Figure \ref{fig:fault_injection_overview}(d). Electromagnetic field causes voltage and current fluctuations inside the device, leading to the faults \cite{dumont2019electromagnetic}.

\color{black}
\textbf{Malicious Hardware:} Malicious hardware is a significant security concern in modern SoC platforms. This attack usually originates by inserting malicious hardware or performing malicious modifications in the original design \cite{tehranipoor2010trustworthy, bhunia2018hardware, mishra2017hardware}. Malicious hardware can be implanted by third-party IP vendors, untrusted design, fabrication, and test facilities. It can potentially violate either of these following.
\begin{itemize}
    \item \textit{Confidentiality: }It can leak a sensitive information to an untrusted observable points. 
    \item \textit{Integrity: } It can illegally alter/modify a data. 
    \item \textit{Availability: }It can affect the availability of the device. 
\end{itemize}

\begin{table*}[]
\caption{IP-level parameters.}
\label{tab:ipLevelParam}
\centering
\begin{tabular}{|l||*{5}{c|}}\hline
\backslashbox{Parameters}{Threat}
&\makebox[3em]{IP Piracy}&\makebox[7.5em]{Power Side-Channel}&\makebox[5.5em]{Fault-Injection}&\makebox[7.5em]{Malicious Hardware}&\makebox[5em]{Supply chain}\\\hline\hline
Locking key size &\checkmark& \textendash &  \textendash &  \textendash & \textendash  \\\hline
Locking key distribution & \checkmark &  \textendash &  \textendash &  \textendash & \textendash \\\hline
Output corruption & \checkmark &  \textendash &  \textendash &  \textendash & \textendash \\\hline
Locking mechanism & \checkmark &  \textendash & \textendash &  \checkmark & \textendash \\\hline
Key error rate (KER) & \checkmark &  \textendash &  \textendash &  \textendash & \textendash \\\hline
Input error rate (IER) & \checkmark &  \textendash &  \textendash &  \textendash & \textendash \\\hline
Mode of operation &  \textendash & \checkmark &  \textendash &  \textendash & \textendash \\\hline
Cryptographic key size &  \textendash & \checkmark &  \textendash &  \textendash & \textendash \\\hline
Number of clock cycles per cryptographic operation &  \textendash & \checkmark &  \textendash &  \textendash & \textendash \\\hline
Data dependent switching activity &  \textendash & \checkmark &  \textendash &  \textendash & \textendash \\\hline
Data independent switching activity &  \textendash & \checkmark &  \textendash &  \textendash & \textendash \\\hline
Gatey type/sizing & \textendash &  \textendash &  \checkmark &  \textendash & \textendash \\\hline
Nearby cells & \textendash &  \checkmark &  \checkmark &  \textendash & \textendash \\\hline
Static probability &  \textendash &  \textendash & \textendash &  \checkmark & \textendash \\\hline
Signal rate &  \textendash &  \textendash & \textendash &  \checkmark & \textendash \\\hline
Toggle rate &  \textendash &  \textendash & \textendash &  \checkmark & \textendash \\\hline
Fan-out &  \textendash &  \textendash & \textendash &  \checkmark & \textendash \\\hline
Immediate fan-in &  \textendash &  \textendash & \textendash &  \checkmark & \textendash \\\hline
Lowest controllability of inputs &  \textendash &  \textendash & \textendash &   \checkmark & \textendash \\\hline
%Average and standard deviation of input controllability &  \textendash &  \textendash & \textendash &  \checkmark & \textendash \\\hline
\end{tabular}
\vspace{-1em}
\end{table*}

\textbf{Supply Chain: }Modern SoC design supply chain comprises several steps, such as defining design specification, RTL implementations, IP integration, verification, synthesis, design-for-test (DFT) and design-for-debug (DFD) structure insertion, physical layout, fabrication, testing, verification, packaging, and distribution \cite{asadizanjani2021counterfeit, stern2019emforced, wang2019system, yang2017cdta, asadizanjani2017counterfeit, asadizanjani2016database, xiao2016circuit, yang2015rfid}. The globalization of these steps makes the IC design increasingly vulnerable to different supply chain attacks. As the design moves through the supply chain, the risk of the design being exposed to different supply chain attacks also increases. Common supply chain attacks that put the hardware design flow at risk are: Recycling, Remarking, and Cloning attacks.
\begin{itemize}
    \item \textbf{Recycling: } As the name suggests, Recycling refers to claiming a used electronic component as a new product. In this type of attack, the attackers extract the chip from a used system. Then with little or no modification, they sell the chip as a new product. The recycled chip usually shows a degraded performance and a shorter lifetime \cite{bhunia2018hardware, asadizanjani2021counterfeit, stern2019emforced, wang2019system, yang2017cdta, asadizanjani2017counterfeit, asadizanjani2016database, xiao2016circuit, yang2015rfid, tehranipoor2015counterfeit, shahbazmohamadi2014advanced, guin2014counterfeit, guin2014counterfeit, guin2014comprehensive, tehranipoor2014tutorial, tehranipoor2014counterfeit, tehranipoor2014counterfeit, guin2013anti, guin2013selection, guin2013counterfeit}.

.
    \item \textbf{Remarking: } Each electronic chip is marked with some information so that it can be uniquely identified. Examples of such information are part identifying number (PIN), lot identification code or date code, device manufacturer’s identification, country of manufacture, electrostatic discharge (ESD) sensitivity identifier, certification mark, etc. \cite{bhunia2018hardware}. Surely, this marking information plays an important role. For example, a space-graded chip can withstand extreme conditions such as a wide range of temperatures and radiation that commercial-graded chips are not capable of tackling. Hence, the space-graded chips cost more. The attacker can remark the commercial-grade chip and sell it in the market at a higher price. In addition, using this remarked chip in spacecraft will surely lead to a disastrous failure. 
    \item \textbf{Cloning: }Attackers can clone a design due to the malicious intent of IP piracy. For example, a dishonest platform integrator can clone the IP and sell it to another platform integrator. 
\end{itemize}

\section{IP level Security Metrics and Design Parameters}
In this section, we present IP-level security metrics for five different threat models. We also discuss the design parameters that contribute to the security metric at the IP level.
%\todo[inline]{add introductory sentences of this section to create the flow}
%\todo[inline]{add IP-level metric where it fits}

\subsection{Metrics to Assess an IP's Vulnerability to Piracy and Reverse Engineering} This subsection briefly describes the existing metrics to evaluate security against IP Piracy. As discussed in Section \ref{sec:threatModel}, logic locking has been a prominent solution to prevent IP piracy by locking the design with a key. However, attackers leverage modern SAT solvers to retrieve the key in a very reasonable time frame. Hence, as a measure of the robustness against IP Piracy, we aim to discuss different metrics that represent the SAT attack resiliency. The most commonly used metrics to evaluate the resistance against IP Piracy are listed below.

\textbf{Output Corruptibility  \cite{shamsi2018approximation}}: The output corruptibility is defined as the probability that the two outputs from a locked IP with wrong key and the functional unlocked chip are not equal for the same input pattern. If $C_e$ and $C_o$ represent the locked and functional IP, respectively, and the input pattern space and wrong key space are represented by $I$ and $K$ respectively, the output corruptibility $C_r$ is represented as follows.

\noindent
$C_r(C_e, C_o)$$\equiv$$P_r[C_e(i,k)\neq C_o(i)], where$  $ i \in I, k \in K$
\noindent
Output corruptibility is a very efficient measure of hiding the design's functionality. However, higher output corruptibility may help SAT solving tools to a quicker convergence.

\textbf{Number of SAT Iterations \cite{xie2016mitigating}:} This is another important metric that represents how many iterations the SAT attacker needs to reveal the key. Usually, the higher is the number, the better the resiliency is. However, the higher iteration number does not necessarily ensure that the key retrieving time is also higher. It depends on different initial conditions and how the SAT solver walks through the DIP search space. For two different designs, one having a higher number of iteration does not ensure it has higher resiliency over the other one. Also the time for each iteration is also not necessarily equal. However, usually a very high number of iteration is expected to take longer in terms of time.

\textbf{CPU Time \cite{morawiecki2013sat}:} CPU time indicates the time needed by the SAT attacking engine to extract the correct key. In this case, timeout is the expectation for an ideal SAT resilient design. The higher the time is, the better is the robustness of the design. 

\subsection{IP Level Parameters Contributing IP Piracy Security Metrics}
The IP-level parameters refer to the different design parameters of an IP at RTL/Gate-level that contribute to quantifying the resiliency against a particular threat. This subsection identifies a set of such parameters that play an important role while evaluating IP Piracy robustness against the SAT attack. Table \ref{tab:ipLevelParam} includes different important IP-level parameters for all the five threats.

\begin{comment}

\begin{table}[]
\caption{IP-level parameters contributing to quantify SAT attack resiliency against IP protection.}
\label{tab:ipLevelParamSAT}
\centering
\begin{tabular}{|l|}
\hline
Key size               \\ \hline
Key distribution       \\ \hline
Output corruption      \\ \hline
Locking mechanism      \\ \hline
Key error rate (KER)    \\ \hline
Input error rate (IER) \\ \hline
\end{tabular}
\vspace{-1.8em}
\end{table}

\end{comment}

\textbf{Locking Key Size:} Key size is one of the essential IP-level parameters that contribute to IP Piracy resiliency. The ideal SAT attack resilient locking mechanism expects to put the attacker in a situation where she cannot retrieve the key without brute force. In practice, the locking mechanism is implemented in a way so that the number of iterations to retrieve the key becomes as close as the number of iterations during brute force. Under this assumption, the increasing key size should increase the SAT resiliency exponentially.  

\textbf{Locking Key Distribution:} Another important IP-level parameter is the distribution of the key. The key distribution represents the fact of whether one locking key is shared among more than one IP or not. In the case of a shared key locking mechanism, retrieving a single key helps to retrieve the functionality of more than one IP. 

\textbf{Output Corruption:} The output corruption represents the deviation of the output of a locked IP from the correct output. As the main idea of locking an IP is to hide the functionality of the design, the deviated output with the wrong key serves this purpose and prevents the attacker from retrieving the functionality of the IP. Usually, the output corruption is measured in terms of the Hamming distance between the correct and the wrong output. Ideally, a 50\% hamming distance is considered the highest deviation. Now, from the SAT attackers’ point of view, output corruption plays a reverse role than the locking point of view. The highly deviated output usually helps SAT solver prune out more number of DIPs in a single iteration. The ideal case for the SAT resilient locking mechanism is that the attacker should be able to prune out only one DIP for one wrong key that eventually leads the attacker to perform a brute force attack.  

\textbf{Locking Mechanism:} The locking mechanism refers to the fact that how the locking is actually implemented in the design. To be specific, it represents how the key gates are placed inside the design, how many fanouts are affected by the key gate, and eventually, what the output corruption is. The researchers have shown that locking with randomly inserted key gates is susceptible to sensitizing the key bit to the output that leads to retrieving the correct key \cite{rajendran2012security}. 

\textbf{Key Error Rate (KER): }$KER$ \cite{liu2021robust} of a key represents the fraction of the input minterms corrupted by the key \cite{liu2021robust}. For an input length $n$ bits, if $X_K$ is the set of the corrupted input minterms by the key $K$, then the $KER$ for key $K$ is defined as

\[KER\ = \dfrac{|X_k|}{2^n}\]
\noindent
Higher value of $KER$ tends to rise the overall output corruptibility of the design.

\textbf{Input Error Rate (IER): }$IER$ \cite{liu2021robust} of an input minterm is represented as the ratio of the number of wrong keys that corrupt the input minterm to the total number of wrong keys \cite{liu2021robust}. For a given input minterm $X$, if it is corrupted by the set of the wrong keys $K_{X}$, and the $K_{WK}$ denotes the set of all wrong keys, then $IER$ is defined as

\[IER = \dfrac{|K_{X}|}{|K_{WK}|}\]
\noindent
$IER$ contributes to the functional corruptibility of the design, hence higher $IER$ helping in hiding the correct function of the design.

\subsection{Metrics to Assess an IP's Vulnerability to Power Side-Channel (PSC) Attacks } 
This subsection discusses some existing security metrics and IP-level design parameters for power side-channel analysis at the IP level.

\textbf{Signal-to-Noise Ratio (SNR):} SNR is the ratio of the variance of actual power consumption to the additive noise \cite{yano2017signal}. If $P_{signal}$ and $P_{noise}$ denote the power consumption of the target attack gates and the additive noise, respectively, the SNR is defined as
\begin{equation}
    SNR = \frac{Var(P_{signal})}{Var(P_{noise})} 
\end{equation}
where var represents the variance of a function.
The signal-to-noise ratio is a measure of the difficulty level to retrieve the correct key by analyzing the power traces.

\textbf{Measurement to Disclose (MTD):} Measurement to disclose (MTD) is defined as the required number of power traces to reveal the correct key successfully \cite{mangard2004hardware}. MTD depends on both the signal-to-noise ratio (SNR) and the correlation coefficient $\rho_{0}$ between the power model and the signal in the power consumption. Mathematically, MTD is defined as 

\begin{equation}
    MTD \propto \frac{1}{SNR*\rho_{0}^2}
\end{equation}

\begin{comment}
\noindent
Although the definition of these metrics differs from one to another, they are based on the collected power traces. The measure of the signals and the noise in the power trace is subject to change while collected from the SoC due to the additional SoC-level parameters discussed in the previous section. All these metrics are efficient in quantifying IP-level PSC robustness. However, none of these considers the SoC-level scenarios.
\end{comment}

\textbf{Test Vector Leakage Assessment (TVLA):} TVLA \cite{gilbert2011testing} utilizes Welch’s t-test to evaluate the side-channel vulnerability. TVLA works on two sets of power traces. One is generated from the fixed key and fixed plain text and another from the same fixed key and random plain text. Then a hypothesis testing is performed on the two sets of traces, assuming a null hypothesis that the two sets of traces are identical. The accepted null hypothesis represents that the collected traces do not leak information about the key. A rejected hypothesis indicates that the traces can be exploited to retrieve sensitive information. 

TVLA is defined as follows
\begin{equation}
    TVLA = \frac{\mu_{r}-\mu_{f}}{\sqrt{\frac{\sigma_r^2}{n_r}+\frac{\sigma_f^2}{n_f}}}
\end{equation}

\noindent
where $\mu_f$ and $\sigma_f$ represent the mean and standard deviation of the set of traces with fixed key and fixed plain texts. $\mu_r$ and $\sigma_r$ are the mean and standard deviation of the set of traces with the same fixed key but random plain texts. $n_r$ and $n_f$ are the numbers of traces in the set of traces with random and fixed plaintext, respectively.

\textbf{Kullback Leibler (KL) Divergence:} KL divergence \cite{kullback1951information} is generally used to measure the statistical distance between two different probability distribution functions. The notion of the statistical distance is leveraged in power side-channel analysis to identify the vulnerable design. Suppose the distribution of the power consumption of a cryptographic algorithm is different for different keys \cite{he2019rtl}. In that case, it indicates that the power consumption can be exploited to reveal information about the key. 

If  $f_{T|k_i} (t)$ and $f_{T|k_j} (t)$ are the two probability density functions of the power consumption for the given keys $k_i$ and $k_j$, respectively, the KL divergence is defined as
\begin{equation}
    D_{KL}(k_i||k_j)=\int {f_{T|k_i}(t) \log{\frac{f_{T|k_i}(t)}{f_{T|k_j}(t)}}dt}
\end{equation}

A larger value of KL divergence implies that the power consumption for different keys differs significantly and that the design can be easily exploited by power side-channel analysis. On the other hand, the lower magnitude of the KL divergence indicates greater robustness against the power side-channel attacks.

\textbf{Success Rate (SR):} Success Rate refers to the ratio of the successful attack to retrieve the correct key to the total number of attacks attempted \cite{standaert2009unified}. It is defined as 

\begin{equation}
    SR = \frac{Number of successful attacks}{Total number of attacks}
\end{equation}

\textbf{Side-Channel Vulnerability (SCV):} The side-channel vulnerability (SCV) metric is functionally comparable to the widely used signal-to-noise ratio (SNR). However, SCV metric can be utilized in formal methods based on information flow tracking (IFT) to assess PSC  vulnerability with a few simulated traces at the pre-silicon design stage \cite{nahiyan2020script}, as opposed to thousands of silicon traces required in SNR metric. It is defined as

\begin{equation}
    SCV = \frac{P_{signal}}{P_{noise}}=\frac{P_{T.hi}-P_{T.hj}}{P_{noise}}
\end{equation}

where $P_{T.hi}$ and $P_{T.hj}$ denotes the average power consumption of the target function when the Hamming Weight (HW) of the output is $hi = HW (Ti)$ and $hj = HW (Tj)$ for ith and jth input patterns, respectively. For the PSC assessment, the difference between $P_{T.hi}$ and $P_{T.hj}$ is estimated as signal power.

\subsection{IP Level Parameters Contributing Power Side-Channel (PSC) Security Metrics}
We explored and extracted the IP-level design parameters which contribute to the IP-level PSC metric. We briefly discuss those parameters as follows.

\begin{comment}
\begin{table}[h]
\caption{IP-level parameters contributing to quantify PSC resiliency.}
\label{tab:ipLevelParamPSC}
\centering
\begin{tabular}{|l|}
\hline
Mode of Operation                                              \\ \hline
Key length                                                     \\ \hline
Clock Cycles required for a cryptographic operation                 \\ \hline
Data dependent switching activity                              \\ \hline
Data independent switching activity                            \\ \hline
\end{tabular}
\end{table}
\end{comment}

\begin{comment}
\begin{itemize}
    \item Modes of Operation
    \item Key length
    \item Number of bytes the cryptographic algorithm operates at a time
    \item Data dependent switching activity
    \item Data independent switching activity
\end{itemize}
\end{comment}

%\todo[inline]{elaborate each parameters}

\textbf{Mode of Operation:} The total Power consumed during a cryptographic operation can greatly influenced by the mode of operation. For example, AES block cipher algorithm can operate in multiple modes such as Electronic Code Book (ECB), Cipher Block Chaining (CBC), Cipher Feedback (CFB), Counter (CTR) mode, etc. PSC resiliency for different modes of operation will vary \cite{aesmodes2014} as the power consumption and noise level will be different due to the additional logic, use of initialization vector or counter, nonce, etc.
 
\textbf{Cryptographic Key Size:} Key size of the cryptographic algorithm is another important factor because it determines the number of round operation in AES algorithm. The rounds of operation for 128, 192, and 256 bit keys are 10, 12, and 16, respectively. The parallel activities will vary depending on the number of rounds, resulting in varying resiliency against side-channel attacks.

\textbf{Number of Clock Cycles per Cryptographic Operation:} When evaluating resiliency against side-channel attacks, the number of clock cycles required to perform an AES operation in the implementation is the most important factor to consider. A loop unrolling architecture may be used in the AES implementation, in which one or more rounds of operations are performed in the same clock cycle. In the simplest case, only one round of the algorithm is implemented as a combinational processing element, and the results of the previous clock cycle are stored in data registers. Multiple data blocks are processed simultaneously in a clock cycle in a pipelined architecture, and the implementation usually requires multiple sets of registers and processing elements. Furthermore, the key expansion can be done at the same time as the round operations, or during the first clock cycle. As a result, the number of cycles needed for the algorithm will have an impact on the parallel operation and noise in the power trace.

\textbf{Data Dependent Switching Activities:} Data dependent switching activities are the transistor activities directly related to the key. The operations in the round that involve the use of key and outputs generated influence the data dependent activities. For example, the first add round key and substitute operations involve the use of the key value that contribute directly to the power consumption. Moreover, the value of the key and the plaintext also affect the data dependent activities.

\textbf{Data Independent Switching Activities:}
Data independent switching activities are the transistor activities that are not correlated to the input key and adds noise to the power trace. For example, the first round in AES operation adds enough confusion and diffusion to the plaintext that the correlation between the key and power will reduce significantly in the later rounds. Moreover, additional logic and circuitry for different implementation and architecture of AES algorithm will provide different amount of data independent switching. Furthermore, the traces include the power consumption from parallel activities in other IPs present in a system. Depending on the design and application specific parameters of the system, the additional activity added to the noise will vary. If no other IP is allowed to operate during an AES operation, it will result in the worst case scenario from a security perspective.

%\color{blue}
\subsection{Metrics to Assess an IP's Vulnerability to Fault Injection Attacks } 

This Section discusses different parameters that are crucial in fault injection attacks, and can impact the overall feasibility of the fault injection attacks, as discussed below.

\textbf{Spatial Controllability}: allows an attacker to target a specific net/gate in the design. Considering the large design size, not all components (nets/gates/registers) are crucial for a successful attack. An attacker tries to inject faults in specific components, if violated, helps his case to exploit and cause integrity and confidentiality violations. Therefore, the higher the spatial controllability of the fault injection attack, the higher is the design's susceptibility. Several parameters can contribute to spatial controllability, i.e., fault method (clock, voltage, laser, etc.), design's timing information (path delays, clock, etc.), library information (cell/gate types). For example, laser and optical fault methods provide more spatial controllability to an attacker to target specific locations on the chip. In contrast, clock and voltage glitching methods can violate multiple paths in the design, thus injecting multiple faults in the design. Similarly, delay distribution of the paths can dominate which registers will be impacted by the clock and voltage glitching.

\textbf{Temporal Controllability}: allows an attacker to control the fault injection time during the design's execution. For example, to effectively perform differential fault analysis on an AES with minimal faults, an attacker would like to inject fault in the eighth round of the execution~\cite{tunstall2011differential,ghalaty2014differential}. However, attacking after the eighth round would require more faults to retrieve the entire key, and before the eighth round would make the differential fault analysis complex, rendering key retrieval futile. Therefore, if an attacker can control the triggering of the clock, voltage, or laser injection, it directly impacts the attacker's capability to control the faults and thus impacts the design's susceptibility against fault injections.

\textbf{Fault Type and Duration}: type of faults, i.e., permanent, semi-permanent, and transient faults, can also impact the fault injection attacks. For example, a laser injects transient faults in the device, but higher laser power can break the silicon, causing permanent faults in the device. Where transient faults cause bit-flips at the gate's output, permanent faults can lead to stuck-at 0/1 at the gate's output. Therefore, Different types of faults require different analyses from the attacker to leak information \cite{zhang2018persistent}. Similarly, fault duration can also impact the design's susceptibility. For example, if a clock glitch is too small or within the slack relaxation of the timing path, the fault may not get latched to impact the design's functionality. Similarly, the laser's duration can impact the number of clock cycles transient effect of laser current would be observed in the design.

\textbf{Fault Propagation}: is required to ensure if the injected faults can propagate to the observable points (for e.g., ciphertext in AES encryption). Faults injected by clock/voltage glitching, laser injection, etc., if not latched to the register, do not go through logical flow, having no impact on the design's execution. Timing information of the paths, laser stimulation period, fault duration, system's clock, and other factors can impact if the faults will be successfully latched or not.

\textbf{Fault Method}: fault injection method can also play a major role while evaluating/developing fault injection security metrics. For example, clock and voltage glitching-based fault injection methods are global in nature. Meaning it's hard to control the fault location, and a single glitch can cause a fault in multiple paths across the design. In contrast, localized fault injection methods such as laser and optical fault injection are local in nature, and an adversary can target specific fault nodes to inject fault. Also, the physical parameters involved with different methods is very different. For example, laser injection requires the backside of chip to be exposed, whereas clock glitching requires access to the system's clock.

The above discussed parameters are crucial to define the fault metrics to determine the design's suceptibility. At the IP level, faults can occur at the data path or the control path. Data path comprises the data flow through the design from register to register via combinational gates. In contrast, a control path consists of control registers controlling the data path flow, e.g., finite state machines (FSMs).  

Security metric to evaluate fault injections in the data paths is challenging, and no security metrics exist to the best of our knowledge. To perform the evaluation, one has to consider the underlying design (e.g., type of crypto) and the threat model. For example, it has been shown that faults injected in the eighth, ninth, and 10th rounds of AES can assist in leaking keys, whereas faults in earlier rounds are more difficult to exploit~\cite{dusart2003differential}. The hypothesis and the mathematical models developed to exploit faults differ across crypto algorithms.

However, faults on the control paths, like FSM, can allow an attacker to bypass the design states. For example, in AES, an attacker can bypass round operations to directly reach the done state, thus, rendering the security from encryption futile. Thus, to measure the overall vulnerability of FSM to fault injection attacks, the vulnerability factor of fault injection ($VF_{FI}$) has been proposed in \cite{nahiyan2016avfsm} as,

\begin{equation}
    VF_{FI} = {PVT(\%),ASF}
\end{equation}

\noindent where,

\begin{equation}
    PVT(\%)= \frac{\sum VT}{\sum T}, ASF = \frac{\sum SF}{\sum VT}
\end{equation}

The metric is composed of two parameters, i.e., PVT(\%) and ASF. PVT(\%) is the ratio of the number of vulnerable transitions ($\sum VT$) to the total number of transitions ($\sum T$). Whereas ASF is the average susceptibility factor ($\sum SF$). The susceptibility factor is defined as,

\begin{equation}
    SF_T = \frac{min(PV)-max(PO)}{avg(P_{FS})}
\end{equation}

\noindent Where min(PV) is the minimum value of delays in path violated, and max(PO) is the maximum value of delays in Paths not violated.
Accordingly, the higher the PVT(\%) and ASF value, the more susceptible the FSM is to fault injection attacks.

\begin{comment}

\subsection{IP level Parameters Contributing Fault Injection Security Metrics} \label{sec:fault_injection_IP}
To the best of our knowledge, there are no security metrics to quantify the pre-silicon standalone IP's resistance against fault injection attacks. However, various parameters can impact the IP's tolerance against various fault injections. We list many of these parameters and their impact on different fault injection methods.

\end{comment}

\subsection{Metrics to Assess an IP's Vulnerability to Malicious Hardware}
We perform an extensive literature review and extract the existing metric to evaluate the Malicious Hardware in a design. We discuss some of these metrics in the rest of this subsection. 
\textbf{\\Controllability and Observability: }Controllability \cite{kantipudi2010controllability} of a component within a design refers to the ability to control the inputs of the component from the design's primary inputs. On the other hand, observability \cite{kantipudi2010controllability} is the ability to observe the output of a component from the primary outputs of the design. The primary inputs of a design are assumed as perfectly controllable, and the primary outputs are perfectly observable. The measurement of the controllability and observability can be normalized between 0 to 1. \bigskip

\textit{Controllability: }Let's consider a component within a design which has input variables $x_1$ to $x_n$ and output variables $z_1$ to $z_n$. If the controllability is represented by $CY$,  then the value of $CY$ for each output of the component can be calculated by- 
    
    \begin{equation}
     CY(Z_j) = CTF \times \frac{1}{n} \sum \limits_{i=1}^n CY(x_i)
    \end{equation}
 \noindent
 Here, $CTF$ is the controllability transfer function of the component and defined as follows.    
 
     \begin{equation}
         CTF \cong \frac{1}{m} (\sum\limits_{j=1}^m 1 - \frac{|N_j(0) - N_j(1)|}{2})
     \end{equation}
\noindent     
Were $N_j(0)$ and $Nj(1)$ are the number of input patterns for which $z_j$ has value from 0 and 1, respectively. 

\bigskip

\textit{Observability: }Observability $OY$ for each input of a component in the design can be calculated as
 
 \begin{equation}
      OY(x_i) = OTF \times \frac{1}{m} \sum \limits_{j=1}^m OY(z_j)
 \end{equation}
\noindent
Where, OTF is the observability transfer function which is the measure of the probability that a fault in the input of a component will propagate to the outputs. If $NS_i$ is the number of input patterns for which $x_i$ causes a change in output, then the $OTF$ is defined as

\begin{equation}
    OTF \cong \frac{1}{n} \sum \limits_{i=1}^n \frac{NS_i}{2^n}
\end{equation}
 
\textbf{}

\begin{comment}
%% This was the original description copied from the dynetics report on metrics. Looking into the source paper of this metric, this description seems not written well. That's why it has been shortened and nadded next to this comment section.
\textbf{Statement Hardness: }Statement hardness \cite{salmani2013analyzing} is the measure of the vulnerability analysis at behavioral level. In a design, where the statement hardness as a lower value, is vulnerable to malicious hardware insertion attack. The metric quantifies- 

\begin{itemize}
    \item Statement hardness for each circuit statement;
    \item Observability for each circuit signal, based on a weighted value range.
\end{itemize}

Circuit control and data flows determine $W$, $L$ and $U$ for each statement in a circuit where, $W$ represents the weight of the value range, and $L$ and $U$ show the lower and upper limits of the value range. The weight of the value range for a statement can be measure as-
    
    \begin{equation}
        W = \frac{U-L+1}{U_o + L_o + 1}
    \end{equation}

Where, $U$ and $L$ are upper and lower limit of a controlling signal at the top of a conditional stack and $U_o$ and $L_o$ are upper and lower limit of a controlling signal.

\end{comment}

\textbf{Statement Hardness: }Statement hardness \cite{salmani2013analyzing} is the measure of the vulnerability analysis of Malicious hardware insertion at the behavioral level of a design. To be specific, it represents the difficulty of executing a statement in the RTL source code. In a design, the statements with a lower value of statement hardness are vulnerable to malicious hardware insertion attacks. An attacker is most likely interested in targeting this area to insert malicious hardware, hoping that it will be activated rarely with certain trigger condition.

\textbf{Hard-to-Detect: }Hard-to-detect is proposed in \cite{tehranipoor2014integrated} to quantify the areas in the gate-level netlist which are susceptible to Malicious hardware insertion. These areas are the common targets of the attacker, as inserting malicious hardware in this area will reduce the probability of being detected during the validation and verification steps.

\textbf{Code Coverage: }Code coverage \cite{nahiyan2017code} is defined as the percentage of the design source code executed during the functional verification. A higher code coverage represents the lower probability of Malicious hardware insertion as it indicates the lower suspicious area in the design. However, it depends on the quality of the testbench used during the functional verification. In \cite{tehranipoor2014integrated}, the authors utilize the code coverage analysis and propose a technique called \textit{Unsued Circuit Identification (UCI)}. It is used to find out the line of codes in the RTL design which are not executed during the functional verification. These lines of code are the potential target of an attacker to insert the malicious hardware.  

\begin{comment}
\textbf{Rareness of nets \cite{saha2016testability}: }  Let us consider, a combinational circuit containing $N$ instances of basic Boolean gates. Given a net $n$ from this circuit, let $p_n^1$ denote the $1-controllability$ of the net (i.e $Pr(n=1)$), and $p_n^0$ = $Pr(n=0)$ = $(1 - p_n^1)$ denote the $0-controllability$.
A net is said to be “rare net” if either $p_n^1 < \theta$ or $p_n^0<\theta$, where $\theta$ implies rareness threshold.

If the total number of simulation test vectors is represented by T, then 
\begin{equation}
    p_n^1 = \frac{Number of Logics - ones occurring at net n}{T}
\end{equation}

\begin{equation}
    p_n^0 = \frac{Number of logic-zeroes occurring at net n}{T}
\end{equation}
\end{comment}

\textbf{Observation Hardness (OH): } Observation hardness (OH) \cite{farzana2021saif} is defined as the percentage of a primary input propagating to an intermediate node. To determine the OH value, a stuck-at-0 or stuck-at-1 fault is injected into a primary input. This fault can be fully detected, potentially detected, or undetected at an intermediate node, representing the intermediate node's dependency on the primary input. The observation hardness of an internal node can be defined as 

\begin{equation}
    OH = \frac{Total No. of Detected Faults}{Total number of faults injected}
\end{equation}

\bigskip
\subsection{IP Level Parameters Contributing Malicious Hardware Security Metrics}In this subsection, we discuss a set of IP-level parameters contributing to the maliciousness of an IP.

\begin{comment}

\begin{table}[h]
\caption{IP-level parameters contributing to quantify maliciousness.}
\label{tab:ipLevelParamMH}
\centering
\begin{tabular}{|l|}
\hline
Static probability                                      \\ \hline
Signal Rate                                             \\ \hline
Toggle Rate                                             \\ \hline
Minimum Toggle Rate of Subsequent Wires                 \\ \hline
Fan-out                                                 \\ \hline
Immediate Fan-in                                        \\ \hline
Lowest Controllability of Inputs                        \\ \hline
Average and Standard Deviation of Input Controllability \\ \hline
\end{tabular}
\end{table}

\end{comment}

% \usepackage{multirow}
\begin{table*}[]
\centering
\caption{IP and platform level design parameters for IP protection, fault injection, power side-channel, malicious hardware, and supply chain.}
\label{tab:ipAndSoC_LevelParam}
\begin{tabular}{|l|l|l|}
\hline
\textbf{Threat}                      & \textbf{IP-level Parameters}        & \textbf{Platform-level Parameters}                 \\ \hline
\multirow{6}{*}{IP Piracy}           & Locking key size                            & Type of platform-level testing architecture   \\ \cline{2-3} 
                                     & Locking key distribution                    & Bypassing Capability                          \\ \cline{2-3} 
                                     & Output corruption                   & Wrapper Locking                               \\ \cline{2-3} 
                                     & Locking mechanism                   & Accessibility to the internal flip flops      \\ \cline{2-3} 
                                     & Key error rate (KER)                & Compression Ratio                             \\ \cline{2-3} 
                                     & Input error rate (IER)              & Scan Obfuscation                              \\ \hline
\multirow{12}{*}{Power Side-Channel} & Mode   of   operation                                   %& IP switching activities                         \\ \cline{2-3} 
                                     %& -                                   
                                     & On-chip voltage regulator                     \\ \cline{2-3} 

& Cryptographic key size                          & Pipeline depth                                \\ \cline{2-3} 
 &
  \begin{tabular}[c]{@{}l@{}}Number of clock cycles for one \\ cryptographic operation\end{tabular} &
  CPU Scheduling \\ \cline{2-3} 
                                     & Data dependent switching activity   & IP-level parallelism                          \\ \cline{2-3} 
                                     & Data independent switching activity & Shared power distribution network (PDN)                                    \\ \cline{2-3} 
                                     
                                     & -                                   & Dynamic voltage and frequency scaling (DVFS)  \\ \cline{2-3} 
                                     & -                                   & Clock jitter                                  \\ \cline{2-3} 
                                     & -                                   & Decoupling capacitance                        \\ \cline{2-3} 
                                     & -                                   & Communication bandwidth
                                     %Glitch
                                     
                        \\ \cline{2-3} 
                                     & -                                   & 
                                     Clock gating 
                                     %Location of the chip in the floor plan                 \\ \hline
                                     \\ \hline
\multirow{2}{*}{Fault-Injection}     & Gate type/sizing                    & Power distribution network                    \\ \cline{2-3} 
                                     & Nearby cells                        & Decoupling capacitance                        \\ \hline
\multirow{6}{*}{Malicious Hardware}  & Static probability                  & IP wrapper / Sandboxing                       \\ \cline{2-3} 
                                     & Signal Rate                         & IP firewall                                   \\ \cline{2-3} 
                                     & Toggle Rate                         & Bus monitor                                   \\ \cline{2-3} 
                                     & Fan-out                             & Security policy                               \\ \cline{2-3} 
 &
  Immediate Fan-in &
  \begin{tabular}[c]{@{}l@{}}Probability of the malicious IP's activation \\ during critical operation\end{tabular} \\ \cline{2-3} 
                                     & Lowest Controllability of Inputs    & Isolation of trusted and untrusted subsystems \\ \hline
\multirow{6}{*}{Supply Chain}        & -                                   & Electronic chip ID (ECID)                     \\ \cline{2-3} 
                                     & -                                   & Physically unclonable function (PUF)          \\ \cline{2-3} 
                                     &                                     & Path delay                                    \\ \cline{2-3} 
                                     & -                                   & Aging                                         \\ \cline{2-3} 
                                     & -                                   & Wear-out                                      \\ \cline{2-3} 
                                     & -                                   & Asset management infrastructure (AMI)         \\ \hline
\end{tabular}
\vspace{-1.5em}
\end{table*}

\textbf{Static Probability: }Static probability of a signal in the design refers to the fraction of time the signal value is expected to be driven at a logic high (1’b1). Ideally, the probability of being at logic high and low should not be skewed. A signal having a significantly higher value of the probability of being logic high indicates that the signal is rarely driven by a logic low, which can be the potential trigger condition of malicious hardware \cite{hoque2018hardware}.

\textbf{Signal Rate: }The signal rate of a wire in the design refers to the number of transitions from logic high to logic low or logic low to logic high at that wire per second \cite{hoque2018hardware}. A wire having lower signal rates can be a potential candidate for the trigger of a malicious circuit in a design.

\textbf{Toggle Rate: }The toggle rate of a signal in the design is defined as the rate at which the signal switches from its previous value \cite{hoque2018hardware}. Attackers usually utilize a signal having a very low toggle rate to implement the trigger condition of the malicious circuit.

\begin{comment}

\textbf{Minimum Toggle Rate of Subsequent Wires: } For an element in the design having multiple fan-outs, the 
Minimum toggle rate  in the immediate fan-outs of a given net.

\end{comment}

\textbf{Fan-out: }Fan-out is an essential feature for malicious hardware insertion attacks. Fan-out of a net is the number of other logic elements the net propagates to. Usually, malicious hardware is inserted to cause a significant impact on the design. In the case of a denial-of-service attack, the attacker would most likely target a payload that impacts a larger part of the design. Thus a net having higher fan-out may be a good choice of an attacker \cite{hoque2018hardware}.

\textbf{Immediate Fan-in: }A large number of immediate fan-in usually indicates that a large function with low entropy has been implemented with rare activation conditions. Hence, the immediate Fan-in becomes an important IP-level parameter of the malicious hardware attack \cite{hoque2018hardware}.

\textbf{Lowest Controllability of Inputs: }Input signals of a group of logic within the design that have a lower impact on the design outputs are considered suspicious nets \cite{waksman2013fanci}. Usually, these type of nets has very low controllability from the primary inputs of the design. The difficulty in controlling these nets makes them a potential candidate for malicious hardware insertion.

\begin{comment}

\textbf{Average and Standard Deviation of Input Controllability: }The distribution of controllability for all the immediate inputs driving a net \cite{hoque2018hardware}.

\end{comment}

%\subsection{Supply Chain}
%\todo[inline]{add metrics and IP-level parameters for Supply-chain}

\subsection{Metrics to Assess an IP's Vulnerabilities to Supply Chain Attacks} As discussed in Section \ref{sec:threatModel}, we consider Recycling, Remarking, and cloning as part of the supply chain attack. This section briefly presents some existing metrics used in assessing these threats.

\textbf{Metrics for Cloning: }Over the years, strong Physically Unclonable Function (PUF) \cite{herder2014physical} has been very effective in detecting cloned chips. The confidence of detecting cloned chips depends on the quality of the PUF, which leads to the PUF quality metrics being a great measure to assess whether a chip is cloned or not with a certain confidence. Major PUF-quality metrics include \textit{uniqueness}, \textit{randomness}, and \textit{reproducibility}.

\textit{Uniqueness: }Uniqueness is the measure of distinctive challenge-response pair (CRP) between chips. Ideally, a PUF referring to one chip should produce a unique CRP that other chips cannot produce, which makes PUF so effective in authenticating a chip while detecting cloned one. Inter-chip hamming distance (inter-HD) is a common measure to mathematically calculate the uniqueness over multiple PUFs \cite{maiti2013systematic}. For $n$ number PUFs, if $k$ is the bit length of a PUF response and $R_i$ and $R_J$ are responses from $PUF_i$ and $PUF_j$, respectively, then inter-HD is calculated as follows.

\begin{equation}
         HD_{inter} = \frac{2}{n(n-1)} \sum\limits_{i=1}^{n-1} \sum\limits_{j=i+1}^{n}  \frac{HD(R_{i}, R_{j})}{k} \times 100\%
     \end{equation}

An inter-HD value of 50\% is considered to be the ideal \textit{uniqueness} as it represents the maximum difference between two PUF responses.

\textit{Randomness: }Randomness indicates the unpredictability of a PUF's response. A PUF can only be used to identify cloned chips when an attacker cannot predict its response. Ideally, a PUF's response is free from all correlations, hence cannot be properly modeled. A PUF with good diffusive property shows good randomness in its response. The diffusive property refers to the fact that a slight change in the input challenge causes a significant variation in the output response. 

\textit{Reproducibility: }Reproducibility measures a PUF's ability to produce the same challenge-response pair in different environmental conditions and over time. Quantitatively, the reproducibility is calculated as the intra-PUF hamming distance as follows \cite{maiti2013systematic}.

\begin{equation}
         HD_{intra} = \frac{1}{m} \sum\limits_{y=1}^{m} \frac{HD(R_i, R^{'}_{i,y})}{k} \times 100\%
     \end{equation}

Here, m is the number of samples for a PUF, k is the length of the response, HD($R_i$, $R^{'}_{i,y}$) represents the hamming distance between the response $R_i$ and the response of $y_{th}$ sample $R^{'}_{i,y}$). 

An ideal PUF is expected to always produce the same response to a particular challenge under different operating conditions with 0\% intra-PUF HD.

\textbf{Metric for Recycling and Remarking: }Guin et al. \cite{guin2014comprehensive} developed a metric called Counterfeit Defect Coverage (CDC) to evaluate the counterfeit detection techniques.  

\textit{Counterfeit Defect Coverage (CDC): }CDC represents the confidence level of detecting a chip as counterfeit after performing a set of tests. It is defined as 

\begin{equation}
    CDC = \frac{\sum \limits_{j=1}^{n} (X_{Rj} \times DF_j)}{\sum \limits_{j=1}{m}} \times 100 \%
\end{equation}

Where $DF_j$ is the defect frequency of the defect j, which indicates how frequently the defect appears in the supply chain. $XR_j$ defines the confidence level of detecting the defect $j$ by a test method $R$. 

\bigskip
\noindent
These three attacks suit well in platform-level discussion, and hence, there is no exhaustive analysis on IP-level parameters so far. We continue our discussion on these three attacks in later sections while discussing the platform-level security for different threats.

\begin{figure*}[t]

% \vspace{-0.5em}
    \centering
    \includegraphics[width=0.7\linewidth]{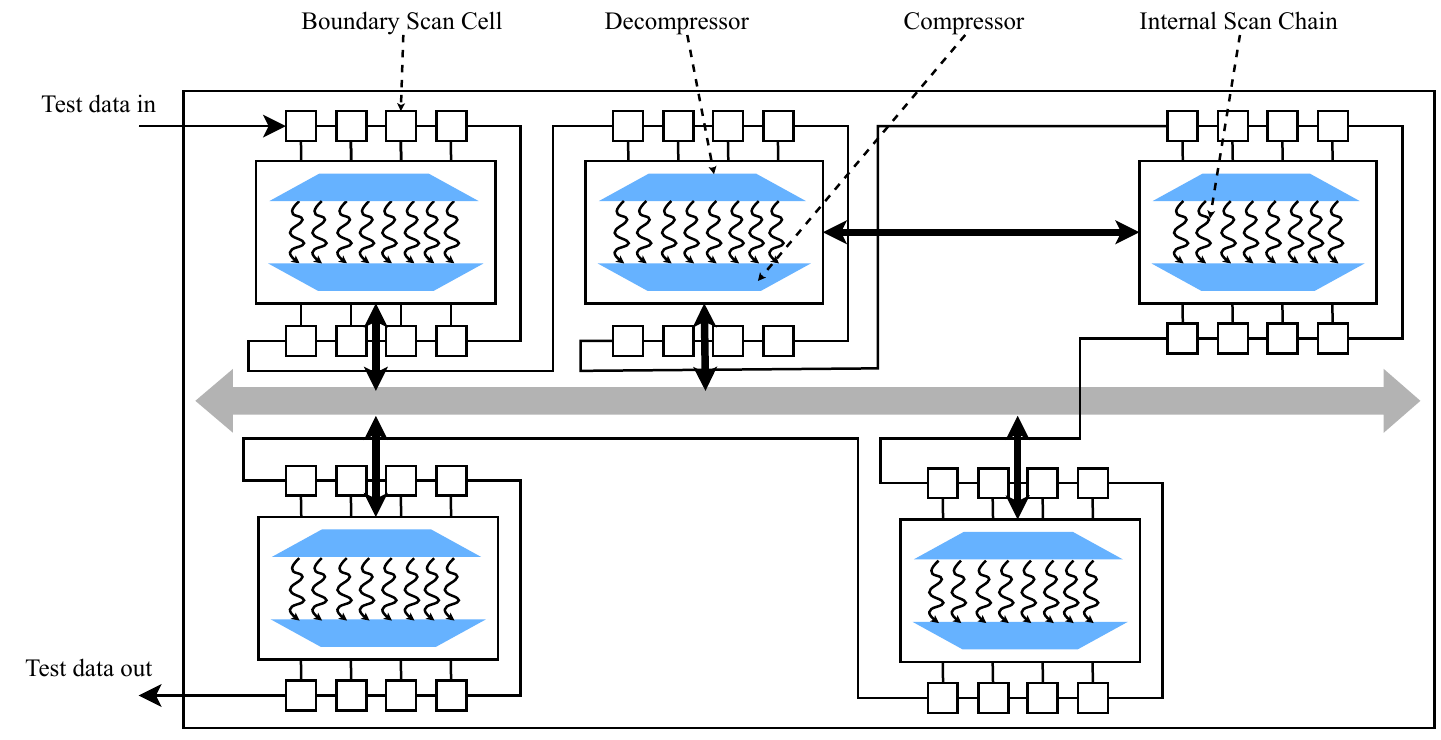}
 %   \vspace{-1em}
  %\vspace{-1.5em}
    \caption{Platform level parameters affecting resiliency against IP piracy.}
    \label{fig:reSOC_LevelParam}
   \vspace{-1.2em}
\end{figure*}

\section{Transition from IP to Platform}\label{sec:ipToSoC_Transition}
In this section, we steer our discussion from IP to platform. So far, we have discussed different IP-level security metrics and design parameters contributing to those security metrics for all five threats. This section focuses on how the design parameters evolve while moving from IP to the platform. In other words, we show different design parameters introduced during the transition from IP to the platform, potentially impacting security metrics at the platform level.

While moving from IP to Platform, an IP gets surrounded by its neighboring IPs. The glue-logic is implemented to connect the IPs, which is not considered while developing metrics at the IP level. A Platform-level testing architecture is introduced that might affect the security at the Platform-level. To get an exhaustive list of such additional parameters, We investigate the transition from IP to Platform and identify several parameters introduced during the integration, which impact the security of the entire platform. We also categorize those parameters based on their impact on a particular threat. We call these additional parameters Platform-level parameters. Table \ref{tab:ipAndSoC_LevelParam} summarizes both the IP- and Platform-level parameters for all five threats. Parameters in the rightmost column of Table \ref{tab:ipAndSoC_LevelParam} are the platform-level parameters that show up after the integration. These parameters are not considered during the assessment of security at the IP level. However, they have a significant impact on security while assessing from the platform point of view. This section briefly discusses all these additional parameters and their impact on IP Piracy, Power Side-Channel Analysis, Fault-Injection, Malicious Hardware, and the Supply Chain attacks at the platform-level.

%\todo[inline]{Add introductory sentences....}
%\subsection{SoC-level Parameters}
\subsection{Platform-level Parameters for IP Piracy }\label{sec:reSocLevelParam} To understand the Platform-level parameters for IP Piracy, let's consider Figure \ref{fig:reSOC_LevelParam}. It shows different Platform-level parameters that affect the SAT resiliency at the Platform level. For example, each IP is wrapped around a wrapper. Compression and decompression circuit have been added to accelerate testing. The wrappers can vary with some of the features such as bypassing capability, access to the internal scan chain, etc. Due to the Platform-level parameters, the way IPs can be attacked while they are standalone significantly differs while they are placed in the Platform. As discussed in Section \ref{sec:threatModel}, SAT attack on a sequential design greatly depends on the access to the internal scan chain. As shown in Figure \ref{fig:reSOC_LevelParam}, the scan access to the target IP inside the Platform is not the same as the standalone IP. Now, the additional boundary scan chain comes with the wrapper. A brief description of these SoC-level parameters affecting the SAT attack resiliency against IP Piracy are discussed below.

\textbf{Type of Platform-level Testing Architecture: }Platform-level SAT resiliency greatly depends on the testing structure of the Platform, as the SAT attack leverages this structure to retrieve the correct key. At the IP level, the retrieval of the key from a locked sequential IP by SAT attack is facilitated by the scan-chain of that IP. Consequently, at the Platform-level, the attack’s success greatly depends on the testing structure of the Platform. Different types of testing structures bring a different level of impact on the IP Piracy resiliency. Example of the Platform-level testing structure includes flattened, direct I/O, concatenated scan chain, etc.

\textbf{Bypassing Capability: }It refers to the capability of bypassing some IPs in a Platform and targeting the specific IP to perform SAT attack. Having bypassing capability in the Platform-level testing structure provides the attacker with the same attacking capability as the IP. However, preventing bypassing capability, although reducing the testing capability, increases the SAT attack's difficulty.

\textbf{Wrapper Locking: }As an industry standard, all the IPs are wrapped around a wrapper during the Platform integration. A locked wrapper boosts the SAT attack resiliency one step ahead as the attacker needs to break the wrapper lock to be able to perform SAT attack at the Platform-level.

\textbf{Accessibility to the Internal Flip-flops: }This is another critical parameter that has a significant impact on platform-level IP Piracy. As discussed above, the bypassing capability provides attackers with the ability to target the specific IP. However, suppose the wrapper is not equipped with the instruction to access the internal flip flops. In that case, the attackers will be left with only the input-output response of that IP through its boundary-scan cells, which makes the attackers’ job a lot more complicated.

\textbf{Compression Ratio: }Performing the SAT attack depends on the controllability and the observability of the internal state of the design through the scan chain. Decompression and compaction introduce the difficulty of precisely controlling internal states and observing the exact response of the internal flip flops. Note that the SAT attack's efficacy depends on the ability to apply the distinguished input patterns and compare the response of both the locked and the functional copy of the IC. The decompressor prevents the attacker from applying the inputs of her choice to the internal flip-flops through the scan chain. On the other hand, the compactor prevents the attacker from comparing the original responses as the shifted output is compacted. Ideally, the attackers need to reverse the compacted output to retrieve the original response before compaction. Usually, the existing compaction circuits used in industries act as a one-way function. It means retrieving the original response from the compacted response introduces some difficulty to the SAT attackers by reducing the observability.

\textbf{Scan Obfuscation: }Scan obfuscation is a great technique to hide the access of the scan chain from the adversaries \cite{rahman2021security, wang2017secure}. With the help of the scan obfuscation, the scanned-in and scanned-out responses are scrambled, and only with the help of the correct key, the original values of the scan chain is retrieved. Hence, it greatly reduces the controllability and the observability of the internal flip-flops in a design. As we already discussed, the SAT attackers’ success greatly depends on the capability of controlling and observing the scan chain circuitry; the scan obfuscation indeed introduces resiliency for the SAT attack at the platform level.

\subsection{Power Side-Channel Analysis}
The Platform-level power side-channel analysis is greatly affected by several parameters that are introduced after the platform integration. Some of these parameters are discussed below.
%\todo[inline]{add introductory sentences...........}

\textbf{Pipeline Depth:} For the sake of simplicity, while we discuss the Power Side-Channel Analysis attack, we consider leaking key from a cryptographic engine by measuring and analyzing the power trace. In the platform-level view, while the cryptographic IP is placed in the platform, it is neighboured by different IPs and different communication buses. %glue logic. 
The Central Processing Unit (CPU) is one of the essential IPs in the platform. Thus, different features associated with the CPU now become the platform-level parameters. The pipeline depth of the CPU is one of the platform-level parameters that indicates the instruction-level parallelism of that CPU. Usually, more parallelism means more activity occurs at the same time. Thus, higher pipeline depth is likely to increase other activity at the same time when the cryptographic IP operates, eventually inducing noise in the power trace of the platform. The increased noise will mask power consumption of the crypto operations and improve resiliency against power side-channel vulnerability. However, this is under the assumption that the instruction-level parallelism will likely increase different IPs activity.

\textbf{CPU Scheduling:} CPU Scheduling indicates the quality of the CPU in performing the parallel operations. The higher CPU utilization indicates a higher probability of the simultaneous activity along with the cryptographic operation, leading to higher noise insertion in the measured power.

\textbf{IP-level Parallelism:} It has been discussed above that the higher pipeline depth potentially increases the instruction-level parallelism. Although the deeper pipeline increases the probability of increasing simultaneous operation, it does not necessarily guarantee that more IPs will operate simultaneously. However, the IP-level parallelism indicates more simultaneous activity from different IPs, which most likely contribute to the noise insertion in the power trace of the platform, leading to the potential increase in the robustness against the Power Side-Channel Analysis attack. One important aspect should be noted that the noise we have been discussing so far is not coming from the total power consumption of the other IPs. Rather the differential power created by the other IPs contributes to the noise. Switching activity from additional IPs and communication buses are the most important factors to contribute to this differential power.

\textbf{Shared Power Distribution Network (PDN):} Shared PDN refers to the fact where different components share the same power rail of the power distribution network (PDN) with the target cryptographic IP. Shared PDN information provides us the notion of an estimated noise that might be inserted in the main rail from where the attackers measure the power. A higher number of IPs sharing the same power rail with the target IP will increase the likelihood of more noise insertion.

\textbf{On-Chip Voltage Regulator:} Modern integrated circuits (ICs) include on-chip voltage regulators to achieve lower noise, better transient response time, and high power efficiency. Such on-chip voltage regulators can be utilized to make the design more resilient against power analysis attacks \cite{yu2017exploiting}. When such techniques are used to scramble or enhance the entropy of the device's power profile, the platform's robustness for PSC vulnerability improves.

\textbf{Dynamic Voltage and Frequency Scaling (DVFS):} Additional noise is used in random dynamic voltage and frequency scaling approaches to randomize power consumption and reduce data-dependent correlation. Such techniques can be employed as a platform-level countermeasure against DPA attack  \cite{baddam2007evaluation}, improving the robustness of the system.

\textbf{Glitch:} Glitching is a common phenomenon in the COMOS circuit. On a high level, glitching refers to the multiple switching of a gate within a single clock cycle. Circuits with higher-order glitches tend to inject more noise in the power grid than the glitch-free design \cite{mangard2005side}. Thus the neighboring IPs having higher-order glitches may help to increase the resiliency of a cryptographic IP at the platform level.

\textbf{Clock Jitter:} An essential step while performing the differential power analysis attack is synchronizing the collected power traces through alignment by the clock edges. However, the clock generator may introduce the clock jitter at the platform level, potentially making the analysis more difficult \cite{clockJitter}. 

\textbf{Communication Bandwidth:} The bandwidth of the communication bus will impact the probable parallel activities in other IPs. A higher bandwidth allows more IPs to be active during crypto operations. The additional switching activities are not correlated to the crypto operations and increases noise in the power trace. As a result, a higher communication bus bandwidth will result in increased resistance against side-channel attacks.

\textbf{Clock Gating:} Clock gating is used in modern circuits to reduce the dynamic power consumption of the clock signal that drives large load capacitance. Power dissipation is minimized in the clock-gated circuits as the registers are shut-off by the control logic during the idle period. Clock-gating techniques can be utilized to improve power side-channel resiliency of a platform. For example, latch-based random clock-gating \cite{tanimura2012lrcg} can obfuscate the power trace with time-shifting approach. Thus, presence of clock-gating logic can affect the power side-channel vulnerability metric of a platform.

\subsection{Fault Injection} 
 In this section, we discuss platform-level parameters that can impact the IPs resistance against fault injection.

\begin{itemize}

    \item \textbf{Nearby IPs:} Nearby IPs to the critical IPs can also have a major impact on the fault resistivity of the cell. For example, a critical IP surrounded by very high switching IPs can diminish the decaps effect, thus increasing the noise margins. This can amplify the effect of voltage and clock glitching attacks.

    \item \textbf{Interconnect:} Interconnects can also have an impact on fault tolerance. A higher density of metal interconnects can make it harder to perform laser fault injection from the front side. Similarly, interconnects length and width can impact the delays, thus impacting timing faults.

    \item \textbf{Power Distribution Networks: }Power and ground rails form the major mesh in the chip, forming many horizontal and vertical loops. Therefore, they are the most susceptible to EM injections. EM injection can cause voltage drops/shoots at both power and ground rails. In addition, both rails can experience drops/shoots at a different rate with respect to the EM pulse power due to asymmetrical couplings \cite{dumont2019electromagnetic}. This voltage swing (i.e., $V_dd-Gnd$) propagates toward the pads while being attenuated along their path. The propagation of the voltage swings results in sampling faults during EM injection.
    
    \item \textbf{Decaps: }Power Grid is the major source of noise in the circuit. Reduced power supply voltages have helped in reducing these noise margins. However, power-hungry circuits (such as ROs) can increase these noise variations, causing delay variations and voltage drops. Decaps help to reduce these noise margins in the supply voltage. Therefore, the distance of the critical registers from Decaps and power grid, supply voltage, the width of power lines, etc., can have a major impact on the susceptibility of the critical register to voltage fault injection. 
    
    \item \textbf{Error Correction and Redundant Circuitry: } Error correction circuits such as bus parities can also impact the design's resiliency against fault injection. For example, parity bits can help resolve single or few bit faults in the data bus depending on the error correction circuit. Similarly, redundant (time or spatial) circuitries in the design can help verify if faults were injected in design \cite{patranabis2015biased}. However, such measures incur heavy area and latency overhead on the design.
\end{itemize}

\color{black}
\subsection{Malicious Hardware}
In this section, we discuss the platform-level parameters that affect the Malicious hardware attack at the platform-level.

\textbf{IP Wrapper: }IP wrapper is usually used at the Platform level, which acts as a checkpoint for the IP not to perform any unauthorized action. Being monitored and controlled by the wrapper, an IP, even including malicious hardware, cannot easily perform the malicious activity. Hence, the IP wrapper impacts the platform-level maliciousness to some degree.

\textbf{IP Firewall: }Like the IP wrapper, an IP firewall monitors and controls the activity of an IP and ensures the expected behavior. The firewall containing a good firewall policy can block an IP's activity from performing malicious action, potentially protecting against the malicious hardware attack at the platform level.

\textbf{Bus Monitor: }A bus monitor is a component that continuously monitors the activity in the BUS connecting different IPs. Monitoring the BUS activity can significantly reduce the malicious data snooping and message flooding with the intent of the denial-of-service attack.

\textbf{Bus Protocol: }Bus protocol is the set of rules that each component in a platform should follow while using the bus. The goal of the Bus protocol is to ensure the correct and secure use of the Bus. A well-developed bus protocol has the potential to reduce the risk of malicious activity during communication through the bus.

\textbf{Security Policy: }Security policies are a set of policies/rules that are enforced by a policy enforcer in a platform. It aims to maintain secure functionality and avoid any potentially unauthorized activity. An exhaustive list of security policies can significantly reduce the malicious activity inside the platform. However, developing the security policy list is a non-trivial task.

\textbf{Probability of the Malicious IP's Activation During Critical Operations: }Not all the IPs are meant to interact with each IP in a platform. Based on the design requirements, each IP has a probabilistic value to interact with a particular IP. The interaction of an IP having a security asset with a malicious IP is more of a concern than the interaction with a non-malicious IP. Hence, the interaction probability becomes an essential parameter in quantifying platform-level security. 

\textbf{Isolation of Trusted and Untrusted Subsystems: }Isolating the untrusted IPs from the security-critical IPs significantly reduces the probability of the occurrence of any malicious activity.

\textbf{Memory Management Unit (MMU) Policy: }Memory is one of the most important components of a platform. MMU policy ensures that no unauthorized application can access the protected memory region, reducing the risk of any malicious activity in the secure memory. 

\subsection{Supply Chain}
The platform-level supply chain attack resiliency is significantly affected by several parameters introduced at the chip level. Some of these parameters are discussed in the rest of this section.

\textbf{Electronic Chip ID (ECID): }ECID is a unique number that is used to tag the chip platform throughout the supply chain. It is used in detecting the remarked chip by tracking its ECID.

\textbf{Physically Unclonable Function (PUF): }Physically unclonable function (PUF) \cite{herder2014physical} utilizes the inherent process variation introduced in every chip during the manufacturing process. This process variation is unpredictable and uncontrollable, making a unique fingerprint of the individual chip used in chip identification and authentication \cite{alkabani2007active,wang2010novel}. PUF is very efficient in detecting the cloning of a platform chip.

\textbf{Path Delay: }In \cite{zhang2012path}, the authors utilize the path delay as the fingerprint to detect the recycled chip. As the recycled chip has been in use for a particular duration of time, it is usual that the performance of the chip will be degraded. Due to the negative/positive bias temperature instability (NBTI/PBTI) and hot carrier injection (HCI), the path delay of a used chip becomes sufficiently larger that it is used to detect whether the chip is recycled or not. The higher the path delay, the higher probability that the chip is recycled.

\textbf{Aging and Wear-out: }Temperature instability (BTI), hot carrier injection (HCI), electromigration (EM), and time-dependent dielectric breakdown (TDDB) are the key parameters for the aging degradation and wear-out of a platform chip \cite{zhang2013design, dogan2014aging, guin2014low, guin2014comprehensive,wang2012representative}. An aged and worn-out chip usually represents a significantly degraded performance compared to the new chip.

\textbf{Asset Management Infrastructure: }The Asset management infrastructure (AMI) is a cloud-based infrastructure that facilitates testing a platform after it is fabricated. It can establish secure communication with the chip on the manufacturing floor. It is also responsible for unlocking and authenticating the chip following the secure protocol, consequently preventing the overproduction of the chip. 

\bigskip
\noindent
All the parameters discussed above contribute to platform-level security to some degree. These parameters either help or hurt the base security (IP-level security) while quantifying the security at the platform level. Section \ref{sec:caseStudy} shows how these parameters are included when the security measurement and estimation is performed for two different threats: IP piracy and Power side-channel analysis.
%\input{methodology}
%\section{Proposed Methodology}
\begin{comment}
\textcolor{blue}{Now discuss how side-channel leakage is impacted in SoC. Just approach. Not the details and simulation results (Leave it for conference papers). Common approach for all threat vectors.
We present the problem and but don’t provide complete solution, rather outline the directions.
}
\end{comment}
\section{Security Measurement \& Estimation}
In previous sections, we started with an IP-level discussion. Then we discussed the transition from IP to the platform and discussed how the parameters evolve during this stage. In this section, we introduce two novel terms called security measurement and estimation. In particular, we discuss both the measurement and estimation from the platform point of view.

\textbf{Security Measurement: }
Security measurement of a design refers to the exhaustive security verification of the design through simulation, emulation, formal verification, etc. In other words, security measurement includes performing actual attacks and analyzing the resiliency of the design against a particular threat. For example, measuring the PSC robustness of a platform requires performing the power side-channel analysis attack on the platform by collecting and analyzing its power traces. If we consider MTD as the security measure for power side-channel analysis, then the number of power traces required to reveal the cryptographic key from the platform would indicate the measured security of the platform. Similarly, measuring security for IP piracy would require performing SAT attack on the platform and measuring the time to reveal the locking key. To measure the maliciousness of platform, an exhaustive security property verification may be a good choice to measure the confidentiality violation through information leakage. This approach can also be used to measure the data integrity violation. PUF, ECID, and age detecting sensors can be used to measure the probability of a chip being recycled, remarked, or cloned.

\textbf{Security Estimation: }Security estimation, on the other hand, does not depend on the simulation or emulation. Instead, it leverages the modeling of the design and different parameters to estimate the security. As a result, the estimation is a lot faster than measurement. Security estimation is a very efficient technique that can enhance the process of building a secure design. For example, with the help of the estimation, the designer can quickly estimate the security of the entire chip earlier in the design phase. Based on this estimated security, the designer can easily make the necessary modification to meet the other PPA (Power, Performance, Area) metric while not compromising the security. As the core component of the estimation, first, a model of the impact factor of the SoC-level parameters is developed. Then this model can predict/estimate the security of an unknown SoC while not needing the actual value of the parameters. Unlike measurement, estimation can be very useful for a larger design. Ideally, the estimation should not depend on the design size. It brings a great benefit to make the designer capable of choosing among different components and different SoC configuration that provides the best result in terms of security. However, the estimated security lacks behind the measured security in terms of accuracy.            

\begin{comment}
which parameters of a chip, it is expected that the measured security is as accurate as possible. Unlike the estimation which is just one sort of prediction of how the security will change while moving from IP to SoC, the measurement should give us the real scenario about the security. The measurement of security of a particular chip includes the exhaustive experimental evaluation. Usually, to get the security measure, a device should be simulated or emulated with the exhaustive workload and all the corner cases that can show up in the field. However, a good selection of workload is a challenge in this regard.
\end{comment}

\begin{comment}
\textcolor{blue}{Clear definition of security estimation and measurement.
Discuss about when the estimation makes sense and when do we need the measurement. For example, for smaller design, the measurement is more appropriate, however, for larger and complex design the measurement will take a long time and the estimation will be more helpful. Scalability. Accuracy and Efficacy}
\end{comment}

%\subsection{Approaches}
\begin{comment}
\textcolor{blue}{The approach should be discussed as global approach not for individual threat. Different threats will be discussed as cases of the approach. 
No methodology will be included. For example, detailed simulation or any results should not be included.  }
\end{comment}

%\textcolor{blue}{Results for test cases: RE, PSC, etc. Rely on small subsystem. Don't need full blown SoC simulation.}

\section{Security Measurement and Estimation Approaches}\label{sec:caseStudy}
In this section, we discuss our proposed estimation and measurement approach for IP piracy and power side-channel analysis attack, as two case studies. We explain each steps of both the estimation and measurement with some example design.

\subsection{IP Piracy}
\subsubsection{Platform Level SAT Resiliency Measurement Flow}
This section provides the step-by-step process for the measurement of platform-level SAT resiliency of an IP. As mentioned in section \ref{sec:ipToSoC_Transition}, the transition from the IP to platform introduces different additional parameters that significantly impact SAT resiliency. Thus, to measure the SAT resiliency at the platform level, we first mimic the platform-level environment by adding the platform-level parameters with the target IP. The first step of performing SAT attack on a design involves converting the design from the gate-level netlist to the conjunctive normal format (CNF). Sometimes this conversion of the design is referred to as SAT modeling. To measure the platform-level SAT resiliency, We first perform the SAT modeling of the design along with the additional platform-level parameters. Then we perform the SAT attack to measure the time the SAT attacking tools take to extract the key at the platform level. For the sake of simplicity in the demonstration, we are considering one platform-level parameter, such as Decompression/compaction. The goal is to measure the SAT attacking time for breaking the lock of an IP while the compression/decompression circuits are associated with that. The measurement is comprised of two steps: Platform-level SAT modeling of the design, and Performing SAT attack. The following subsections present these two steps in detail.

\begin{figure}[t]

 %\vspace{-1.5em}
    \centering
    \includegraphics[width=\columnwidth]{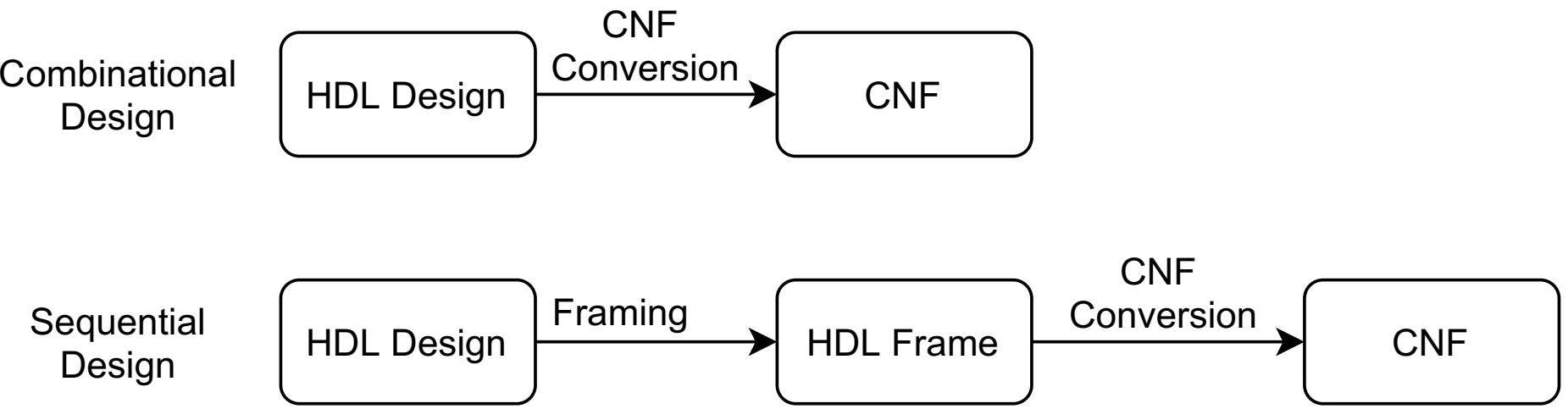}
 %   \vspace{-1em}
  %\vspace{-1.5em}
    \caption{Modeling of the HDL design for SAT attack.}
    \label{fig:combAndSeqModeling}
   \vspace{-1.2em}
\end{figure}

\bigskip
\noindent
\textbf{SAT Modeling for SoC-level Attack: }The existing SAT attacking tools are not capable of performing attack directly on the HDL code. The design needs to be converted into the Conjunctive Normal Form (CNF) before feeding to the SAT attacking tool. The CNF conversion flow for both the combination and sequential design is shown in Figure \ref{fig:combAndSeqModeling}.

\textit{Modeling of Combinational Designs: }The netlist of the combinational designs can readily be converted to the CNF form. There are open-source synthesis tools such abc \cite{abcTool} that can covert the HDL design of a combination circuit into the CNF form.

\textit{Modeling of Sequential Designs: }To understand the Platform-level modeling with the Platform-level parameters, we should first have a solid understanding of IP-level modeling. This section describes the modeling of the sequential design for performing SAT attacks at both IP- and Platform-level.

\begin{itemize}

\item \textit{IP-level SAT Modeling: }The conversion of the sequential designs into CNF format requires an additional step named “framing.” The existing SAT attacking tools has a limitation of performing SAT attack on sequential designs. To address this issue, the sequential design needs to go through the framing process. Framing converts the sequential design into a one-cycle equivalent design. In other words, framing converts the sequential design to equivalent combinational design. For example, we want to perform SAT attack on an AES design that takes 11 clock cycles to complete one cryptographic operation. By framing the AES implementation, we can break the design into 11 pieces of one-cycle design. To get the exact behavior of the sequential design, the 11 frames can be stitched together. However, to perform the SAT attack for measuring SAT resiliency, one frame is sufficient. It is because

\begin{figure}[t]

 %\vspace{-1em}
    \centering
    \includegraphics[width=0.95\columnwidth]{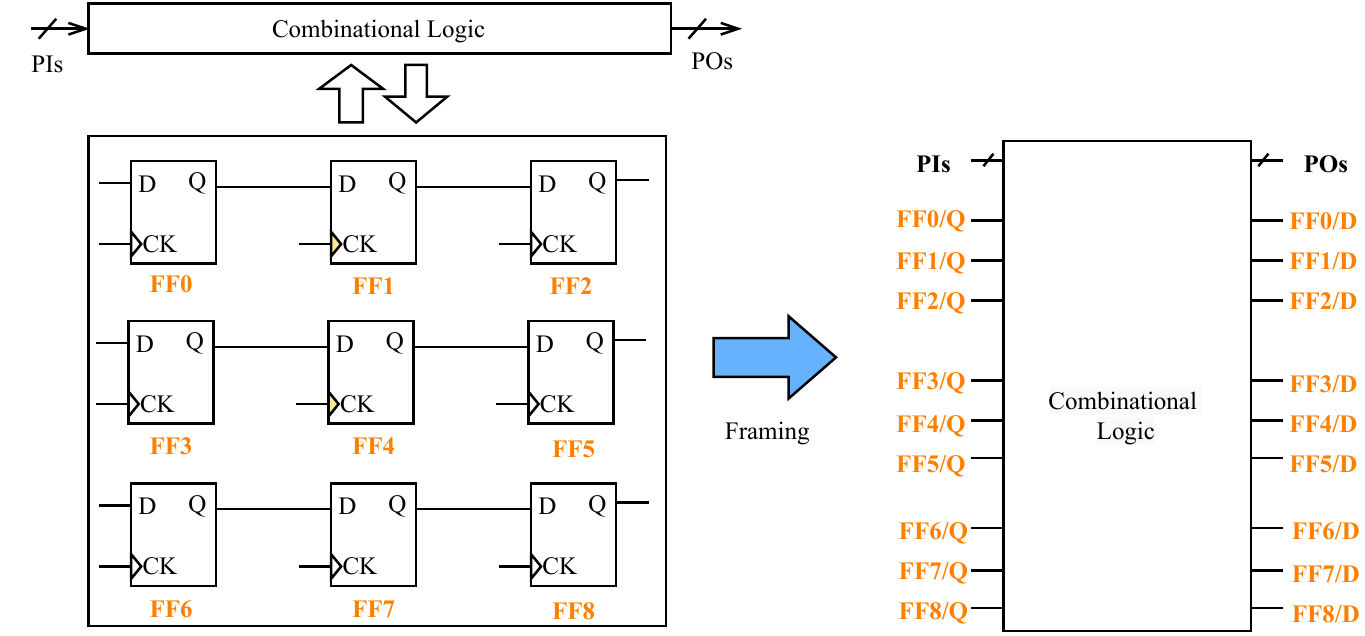}
 %   \vspace{-1em}
  %\vspace{-1.5em}
    \caption{Framing of sequential design.}
    \label{fig:framingSeqDesign}
   \vspace{-1.5em}
\end{figure}

\begin{figure}[b]
 \vspace{-1.5em}
    \centering
    \includegraphics[width=0.8\columnwidth]{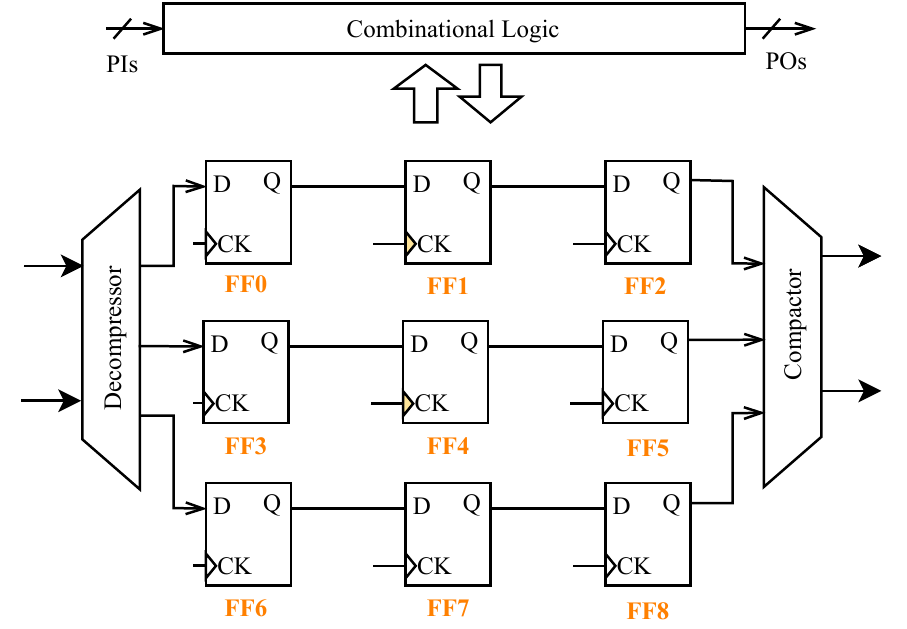}
 %   \vspace{-1em}
  %\vspace{-1.5em}
    \caption{Sequential design with the decompressor and compressor.}
    \label{fig:seqDesignWithCompDecomp}
   %\vspace{-2em}
\end{figure}

\begin{figure*}[]

% \vspace{-0.5em}
    \centering
    \includegraphics[width=0.8\linewidth]{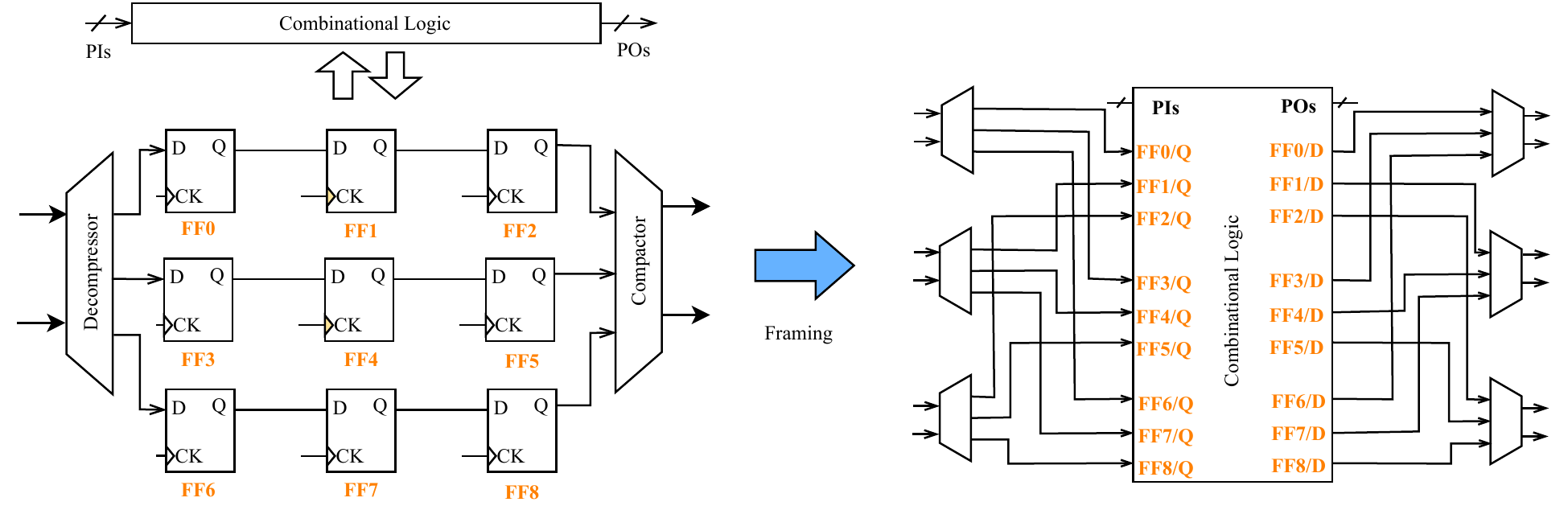}
 %   \vspace{-1em}
  %\vspace{-1.5em}
    \caption{Modeling sequential design with compactor.}
    \label{fig:modelingSeqDesignWithComp}
   \vspace{-1.2em}
\end{figure*}

the SAT attack proceeds by comparing the output of the locked and the unlocked design and pruning the wrong keys away from the search space. For the comparison of the outputs, only one frame is adequate. However, both the locked and unlocked designs should be framed. One question that arises here is, how can one frame the physical chip that is used as an oracle. Here, the scan access to each flip-flop does the same job as framing. Shifting the value through the scan chain, running the design for one clock cycle in functional mode, and shifting out all the flip-flops value gives the same output as one frame would provide. An example of framing of the sequential design is shown in Figure \ref{fig:framingSeqDesign}. The sequential design is the combination of the sequential (flip-flops) and combinational parts. During the framing, all flip-flops are removed from the design, and the corresponding Q pins and D pins are added as primary inputs and outputs, respectively, while the connection between the D and Q pins with the combinational parts are maintained. This frame is then converted into the CNF format.

\item \textit{Platform-level SAT Modeling: }Modeling of the design for the platform-level SAT attack needs to take the platform-level parameters into consideration. In this example of measurement flow, we consider only the impact of the compression ratio (CR) as this is the most important factor in measuring platform-level SAT resiliency. Figure \ref{fig:seqDesignWithCompDecomp} shows a typical sequential design with a decompressor and compactor circuit. Consider the design in Figure \ref{fig:seqDesignWithCompDecomp}, which has three scan chains with three flip-flops in each chain. While applying the desired input pattern, it needs to go through the decompressor, and after three clock cycles, the pattern is shifted into all nine flip-flops. Similarly, while capturing the values out of the flip-flops, they pass through the compactor circuits, and it takes three clock cycles to shift out all the flip-flops’ values. To model the design with the decompressor and the compactor, we need to mimic the exact same behavior in the framed design. Note that the framed design is a one-cycle design. So to mimic the shifting in and shifting out of the data in 3 cycles, three copies of the decompressor and three copies of the compactors should be instantiated instead of one copy. For example, let’s consider the SAT modeling with the compactor circuit. While shifting out the values, FF2, FF5, and FF8 are applied to the compactor in the first clock cycle. Similarly, the group of FF1, FF4, and FF7, and the group of FF0, FF3, and FF6 are applied to the compactor at the 2nd and 3rd clock cycle, respectively. This phenomenon reflects the scenario of resource sharing. In other words, one compactor provides the compression to three groups of flip-flops at different times. Now, if we are interested in shifting out the values of the three groups in one clock cycle, one possible solution would be to connect a separate compactor to each of these groups. In other words, if we connect three compactors to these three groups, we can shift out all the values in one cycle. We leveraged this while modeling the design with the decompressor and the compactor circuit. Figure \ref{fig:modelingSeqDesignWithComp} shows our approach to model a framed design with the compactor circuit. In a framed design, all the flip-flops’ D and Q pins are introduced as primary outputs and primary inputs, respectively. This makes the connection of the compactor even easier. We instantiate three copies of the same compactor and assign each of those into each group of flip-flops. We can see that the one compactor is connected to the group of FF2, FF5, and FF8. Similarly, two other copies of the compactor are connected to the group of FF1, FF4, FF7 and FF0, FF3, FF6. This connection can be made both in Verilog and bench format. In our case, we first converted the design without the decompressor and the compressor into the bench format (CNF). Then we connected the bench format of the decompressor and the compactor with the design bench.

\end{itemize}

% \usepackage{multirow}
\begin{table*}[]
\caption{Summary of experimental setup.}
\label{tab:expSetup}
\centering
\begin{tabular}{|c|c|c|c|c|c|}
\hline
Category                & Benchmark & Number of Gates & Compression Ratio & Number of  keys & Number of experiments \\ \hline
\multirow{4}{*}{Small}  & c499      & 212             & 1,2,4,8,16        & 4               & 20                    \\ \cline{2-6} 
                        & c880      & 404             & 1,2,4,8           & 4               & 16                    \\ \cline{2-6} 
                        & c1355     & 574             & 1,2,4,8,16        & 3               & 15                    \\ \cline{2-6} 
                        & c1908     & 925             & 1,2,4,8           & 4               & 16                    \\ \hline
\multirow{3}{*}{Medium} & k2        & 1908            & 1,2,4,8,16        & 3               & 15                    \\ \cline{2-6} 
                        & c3540     & 1754            & 1,2,4,8           & 3               & 12                    \\ \cline{2-6} 
                        & c5315     & 2427            & 1,2,4,8,16        & 3               & 15                    \\ \hline
\multirow{3}{*}{Large}  & seq       & 3697            & 1,2,4,8,16        & 3               & 15                    \\ \cline{2-6} 
                        & c7552     & 3695            & 1,2,4,8,16        & 2               & 10                    \\ \cline{2-6} 
                        & apex4     & 5628            & 1,2,4,            & 4               & 2                     \\ \hline
\multicolumn{5}{|c|}{Total}                                                                 & 146                   \\ \hline
\end{tabular}
\vspace{-1.5em}
\end{table*}

\begin{figure}[b]

 \vspace{-1em}
    \centering
    \includegraphics[width=\columnwidth]{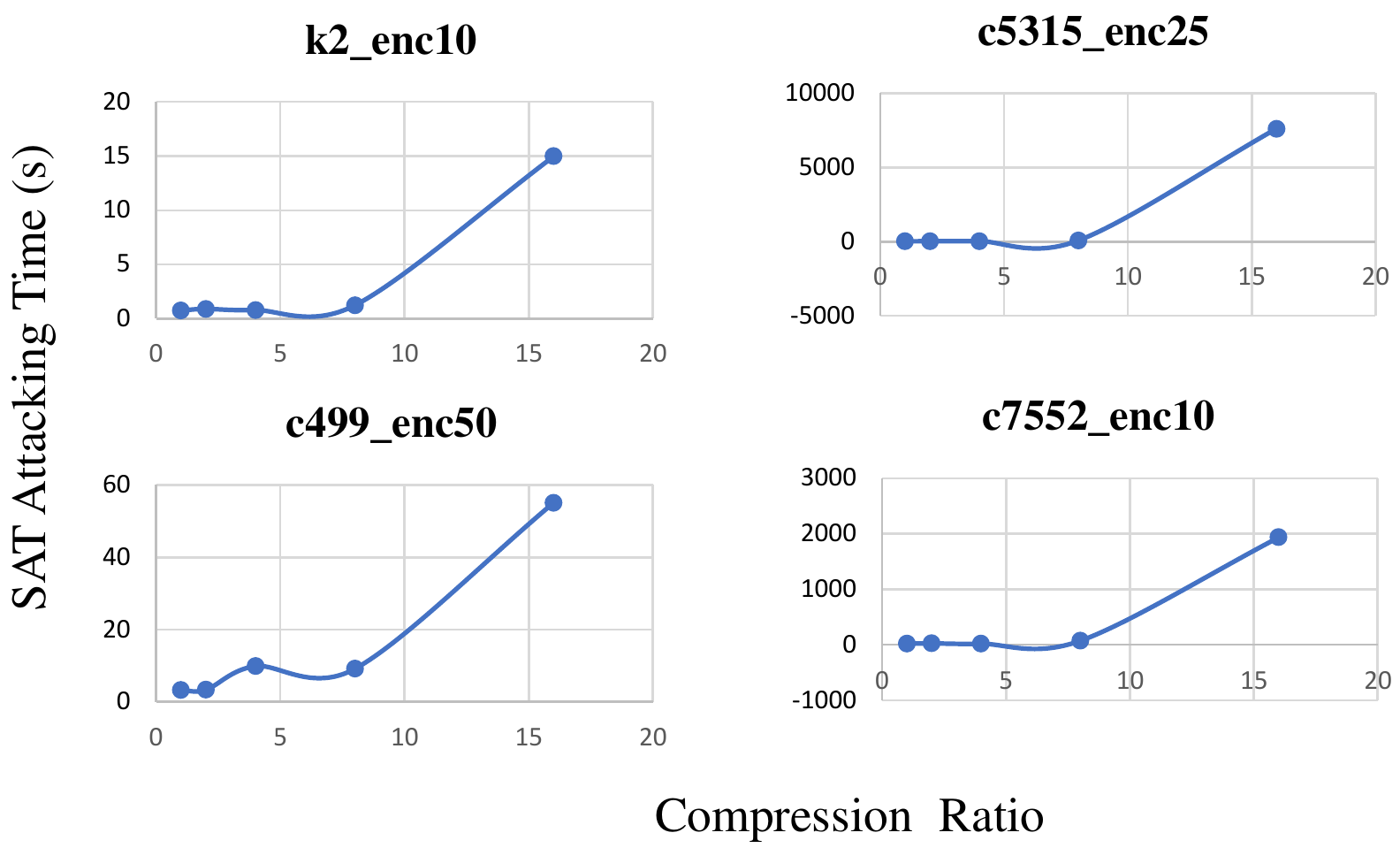}
 %   \vspace{-1em}
  %\vspace{-1.5em}
    \caption{Dominant pattern of CPU time vs CR for four different benchmarks.}
    \label{fig:dominantPattern}
   %\vspace{-1.5em}
\end{figure}

\noindent
\textbf{Performing SAT Attack: }The next step of the platform-level SAT resiliency measurement is performing the actual attack. For this purpose, we used an open-source SAT tool named Pramod \cite{spramod}. We followed the modeling approach discussed above for both the locked and oracle design with the compression and decompression circuit. Then both of this model is fed to the spammed tool to measure the number of iteration and the CPU time to extract the key.

\subsubsection{Platform-level SAT Resiliency Estimation Flow}\label{sec:reEstimation}We developed a data-driven estimation methodology for estimating platform-level SAT attacking time while the IP-level SAT attacking time and the value of the SoC-level parameters (In this case, compression ratio (CR)) are given. Towards this goal, we first built a large data set by measuring the platform-level SAT attacking time for different IPs with different combinations of Key-length and the value of compression ratio. Analyzing the results, we found an interesting relationship between the compression ratio and SAT attacking time. We observe that nearly all the platform-level SAT attacking time for all the IPs follows a general trend with the increasing value of compression ratio. We leveraged this behavior to build our estimation model, which is later used to predict the SAT attacking time for an unknown IP at the platform level.

\textbf{Data Collection and Modeling: }We modeled and performed SAT attack for ten different benchmarks with different values of the compression ratio. We divided the benchmarks into three categories: small, medium, and large. We categorized the benchmarks in terms of the number of gates. The benchmarks which have a number of gates less than 1000 are defined as the small benchmark. Benchmarks containing 1000 to 2500 gates are considered a medium, and the benchmarks having more than 2500 gates are defined as the large benchmark. Each benchmark is locked with a random locking mechanism with four different key lengths. Each version of the design is modeled with the decompressor and compactor of the compression ratio of up to four different values (1, 2, 4, 8, 16). With all these combinations, we performed a total of 146 experiments. A summary of the experiment setup is shown in Table \ref{tab:expSetup}.

\begin{comment}
\begin{table*}
    \centering
    \caption{Summary of the experiments.}
%    \vspace{-1em}
    \label{fig:expSetup}
    \includegraphics[width=\textwidth]{figures/tableExpSetup.png}
\end{table*}

\end{comment}

\begin{figure}[b]

 \vspace{-1em}
    \centering
    \includegraphics[width=0.9\linewidth]{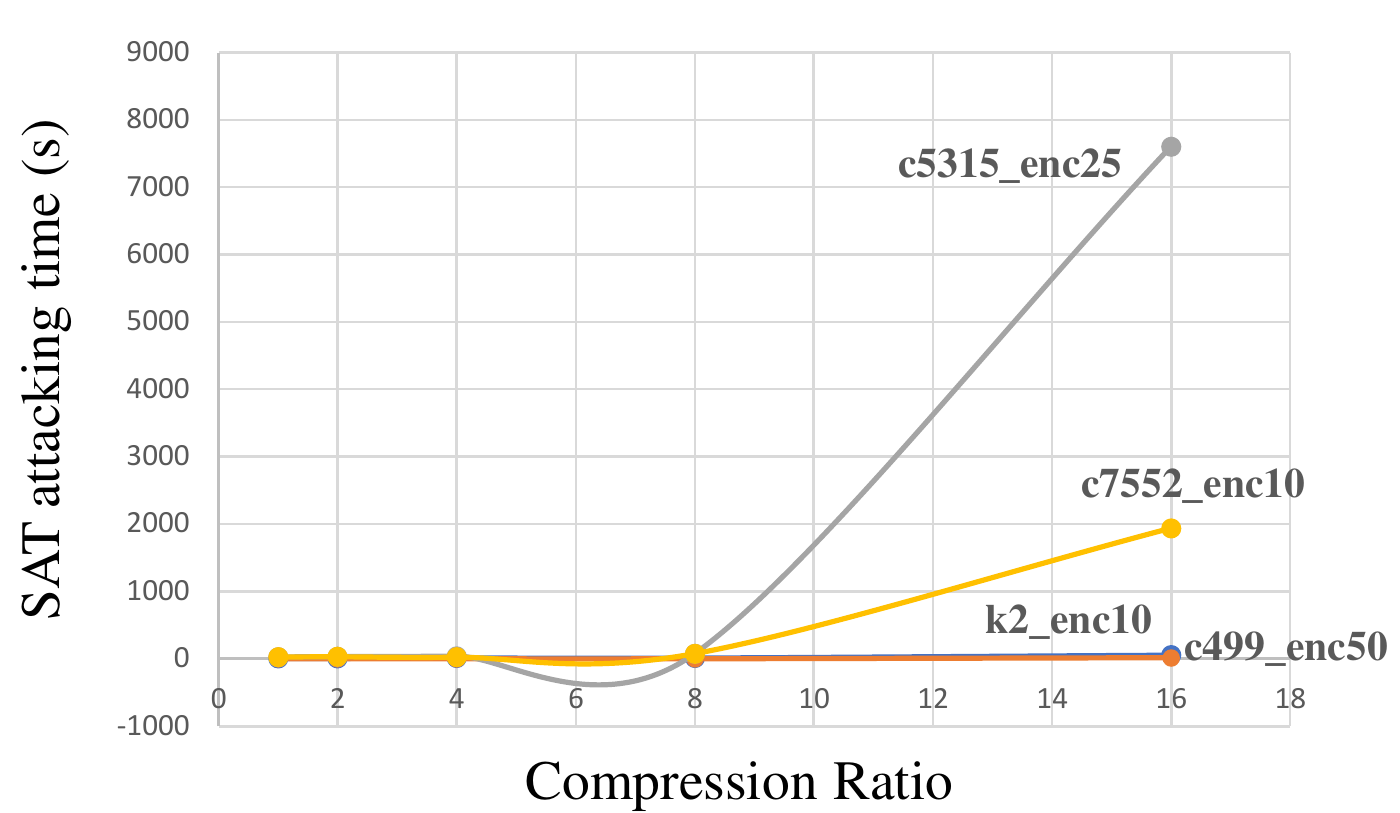}
 %   \vspace{-1em}
  %\vspace{-1.5em}
    \caption{Dominant pattern of CPU time vs CR for four different benchmarks on same scale}
    \label{fig:dominantPatternOnSameScale}
   %\vspace{-1.5em}
\end{figure}

From each of the experiments, we extracted the following features and their values: Key Length, number of gates, number of primary inputs, number of primary outputs, number of scan equivalent I/O (FF IO), Compression ratio, CPU Time, number of total inputs, number of total outputs, and the number of iterations. Analyzing the results, we notice a dominant pattern that exhibits the characteristics of most of the designs. Figure \ref{fig:dominantPattern} shows the dominant pattern of four different benchmarks: c499\_enc50, k2\_enc10, c5315\_enc25, and c7552\_enc10. Basically, this figure shows how the SAT attacking time (in the Y-axis) changes with the increase of the compression ratio (shown in the X-axis). The SAT attacking time at compression ratio 1 is equivalent to the IP-level SAT attacking time. The other compression ratio values mimic the platform-level scenario having the decompressor/compactor with the corresponding compression ratio. As an example plot for k2\_enc10 benchmark, SAT attacking time at compression ratio 1 is around 1 second. This indicates the IP-level scenario. In other words, if the k2\_enc10 was attacked as a standalone IP, the time to retrieve the key would have been around 1 seconds. However, let's consider this IP is placed in a platform where the decompressor and compressor circuit with compression ratio 16 is implemented. In the plot, we can see that the SAT attacking time now increases to 15 seconds (the last point in the plot). This indicates that the IP-level SAT attacking time does not remain the same at the platform level. In this case, higher compression ratio helps to increase the SAT resiliency at the platform level. Similarly, the impact of other platform-level parameters should also be investigated. However, this study is out of the scope of this paper.

The next step is how to build a model with the collected data to estimate the SAT resiliency of an unknown IP at the platform level. The dominant patterns shown in Figure \ref{fig:dominantPattern} seem to follow the same non-linearity. However, when they are plotted on the same scale, their rising rate seems to be different. Figure \ref{fig:dominantPatternOnSameScale} shows the plot of the dominant patterns on the same scale. From this figure, it is obvious that one single model cannot be used to estimate/predict the SAT attacking time for an unknown IP at the platform level. To address this issue, We carefully selected 20 experimental data out of the total 146 experiments to use in the security estimation modeling. While choosing those 20 experimental data, we focus on bringing the possible diversity. These 20 experimental data is the training data set of the estimation model based on which the SAT resiliency of unknown IPs at platform-level is estimated.

\begin{figure}[t]
    %\vspace{-1.2em}
     \centering
     \begin{subfigure}[b]{\columnwidth}
         \centering
         \includegraphics[width=0.6\columnwidth]{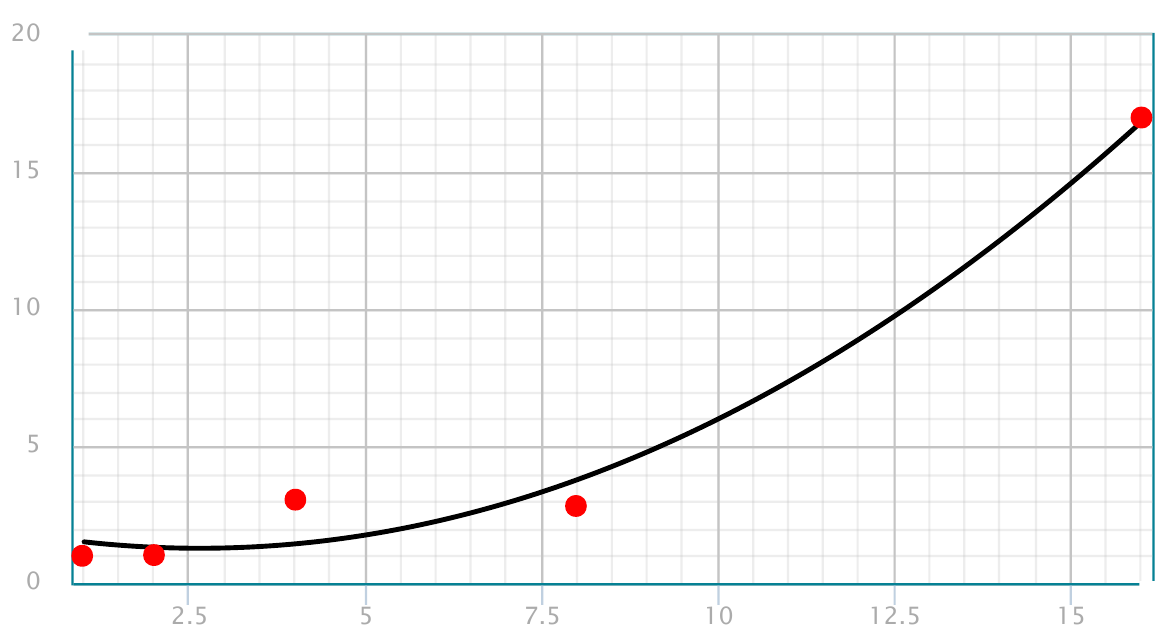}
         \caption{Curve fitting with 2nd order polynomial.}
         \label{fig:three sin x}
     \end{subfigure}
     \hfill
     \begin{subfigure}[b]{\columnwidth}
         \centering
         \includegraphics[width=0.6\columnwidth]{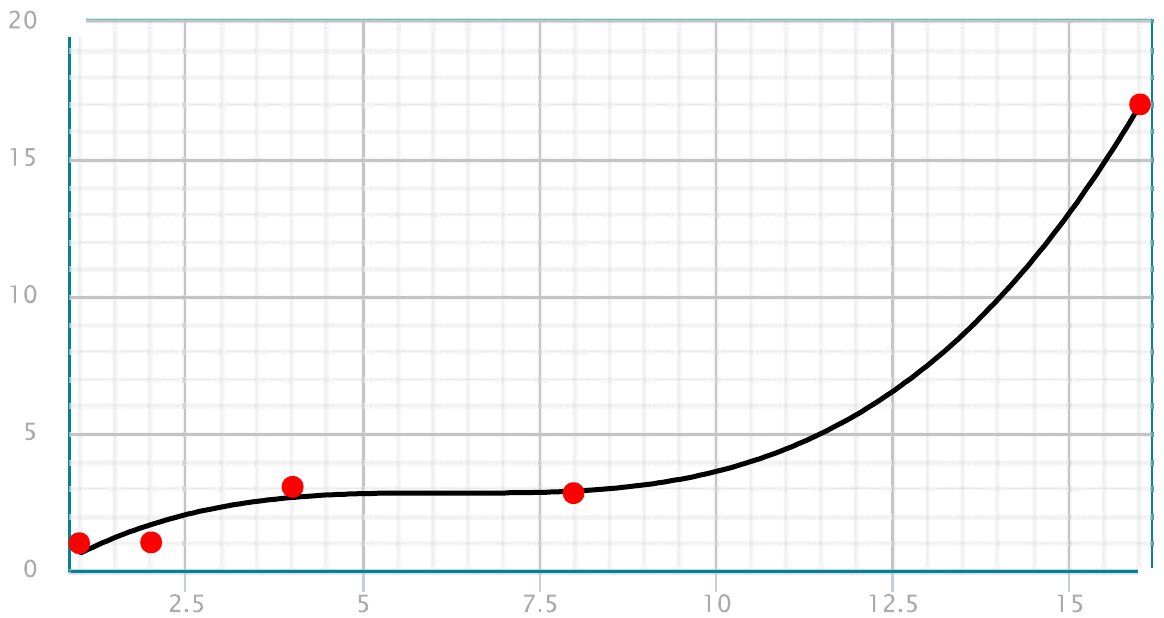}
         \caption{Curve fitting with 3rd order polynomial.}
         \label{fig:five over x}
     \end{subfigure}
        \caption{Curve fitting for the experimental data of c499\_enc50.}
        \label{fig:curveFitting2nd3rdPoly}
\vspace{-1.5em}
\end{figure}

The estimation process comprises two steps: model selection and estimation. As discussed above, the estimation model is built with 20 submodels. While estimating the SAT resiliency for an unknown design, first, it should be identified which submodel (one of the 20 submodels) best suits the unknown design. Towards this goal, we leverage some IP-level metadata of a locked IP. The metadata consists of the following items.
\begin{itemize}
\item Key Length
\item Number of gates
\item Number of Primary Inputs
\item Number of Primary Outputs
\item Number of flip-flop or equivalent ports
\end{itemize}

We utilize the cosine similarity metric for the best match of the metadata of the unknown design with the sub-model. The cosine similarity metric is usually used to find out the angular distance between two vectors. In our case, the metadata is treated as the vector of the items. Then the cosine similarity metric is calculated between the metadata of unknown design and all the 20 sub-models. The sub-model with the highest similarity metric is selected as the estimation model.

\begin{table}[b]
\vspace{-1em}
\caption{SAT attacking time vs compression ratio 1 to 16 for c499\_enc50 benchmark.}
\label{tab:tableForCurveFitting}
\centering
\begin{tabular}{|c|c|}
\hline
Compression Ratio & SAT attacking time (s) \\ \hline
1                 & 1                      \\ \hline
2                 & 1.028257               \\ \hline
4                 & 3.0492296              \\ \hline
8                 & 2.8186724              \\ \hline
16                & 16.9930236             \\ \hline
\end{tabular}
%\vspace{-1.5em}
\end{table}

\begin{comment}
\begin{table}
    \centering
    \caption{CPU time and log10(CPU time) vs compression ratio 1 to 16 for c499 benchmark.}
%    \vspace{-1em}
    \label{fig:tableForCurveFitting}
    \includegraphics[width=0.7\columnwidth]{figures/tableForCurveFitting.png}
\end{table}

\end{comment}

We applied the curve fitting technique to all the 20 experimental data to build the submodels. As an example, Figure \ref{fig:curveFitting2nd3rdPoly} shows the curve fitting approach of one dataset for c499\_enc50 benchmark in Table \ref{tab:tableForCurveFitting} with 2nd and 3rd order polynomial. We choose the 2nd order polynomial over the 3rd order for modeling the relationship between SAT attacking time and the compression ratio to avoid any possible over fitting.

\begin{comment}

\begin{figure}
     \centering
     \begin{subfigure}[b]{0.3\textwidth}
         \centering
         \includegraphics[width=\textwidth]{figures/c499_enc50_Scaled2ndOrderPoly.pdf}
         \caption{Curve fitting with 2nd order polynomial.}
         \label{fig:three sin x}
     \end{subfigure}
     \hfill
     \begin{subfigure}[b]{0.3\textwidth}
         \centering
         \includegraphics[width=\textwidth]{figures/c499_enc50_Scaled3rdOrderPoly.pdf}
         \caption{Curve fitting with 3rd order polynomial.}
         \label{fig:five over x}
     \end{subfigure}
        \caption{Curve fitting for the experimental data of c499\_enc50.}
        \label{fig:three graphs}
\end{figure}

\end{comment}

\begin{comment}

\begin{figure*}[]

% \vspace{-0.5em}
    \centering
    \includegraphics[width=0.7\linewidth]{figures/curveFitting2nd3rdPoly.png}
 %   \vspace{-1em}
  %\vspace{-1.5em}
    \caption{Curve fitting with 2nd and 3rd order polynomial.}
    \label{fig:curveFitting2nd3rdPoly}
   %\vspace{-1.5em}
\end{figure*}

\end{comment}

\begin{comment}

\begin{figure}[]

% \vspace{-0.5em}
    \centering
    \includegraphics[width=0.7\linewidth]{figures/2ndOrderPolynomial.png}
 %   \vspace{-1em}
  %\vspace{-1.5em}
    \caption{2nd order polynomial and the coefficients for estimating the minimum SAT attacking time.}
    \label{fig:2ndOrderPolynomial}
   \vspace{-1.5em}
\end{figure}

\end{comment}

 \begin{figure*}[t]
 \vspace{0.5em}
    \centering
    \includegraphics[width=0.8\textwidth]{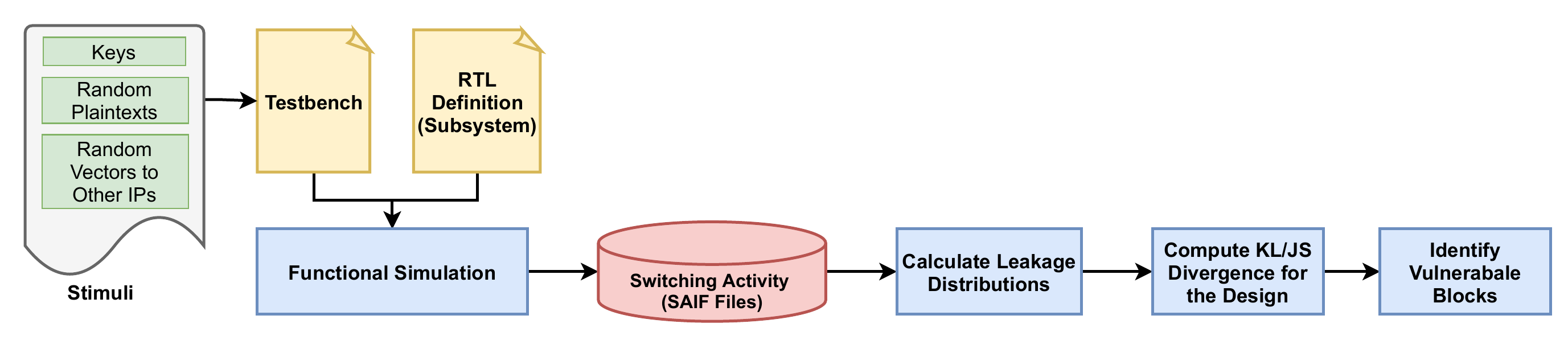}
 %   \vspace{-1em}
  %\vspace{-1.5em}
    \caption{PSC vulnerability measurement workflow.}
    \label{fig:psc_methodology}
   \vspace{-1.5em}
\end{figure*}

\begin{comment}

The design used fo minimum estimation shows the slower increase in the SAT attacking time compared to the increasing value of the compression ratio while the the design for maximum estimation shows faster increase of the SAT attacking time for higher compression ratio. 

\end{comment}

\subsection{Power Side-Channel Analysis}
%\subsubsection{Power Side-Channel (PSC) Vulnerability Assessment}
The PSC measurement and estimation approach focuses on evaluating the PSC vulnerability at the earliest pre-silicon design stage, i.e., register-transfer level (RTL). PSC vulnerability measurement is performed based on the fine-grained simulation of the rtl design. This provides a more accurate approximation of power consumption in each clock cycle of the encryption process. On the other hand, PSC vulnerability estimation depends on a very rough estimate of power consumption based on a few IP attributes rather than an actual simulation of the switching activities. This gives a quick estimate of PSC robustness, but it does not reflect the precise resiliency that the measurement provides. It's worth noting that we limited the scope of the demonstration to see how the most important platform-level factor, i.e. IP-level parallelism, influences power side-channel vulnerability.

\subsubsection{PSC Vulnerability Measurement Flow}

The framework includes two main parts- RTL Switching Activity Interchange Format (SAIF) file generation and identification of vulnerable designs and blocks based on the vulnerability metrics. Then the vulnerable blocks in the design that are leaking information the most are identified for further processing.

\textbf{Subsystem Definition:} In order to analyze the influence of extra noises injected into the power trace by concurrently active non-crypto IPs, such as CPUs or peripherals, we consider a small subsystem for preliminary measurements. Figure \ref{fig:ipToSoC} depicts such a subsystem built from standalone IPs, which includes a sample AES core and a few other IP blocks that execute random logic operations unrelated to cryptography. The AES encryption key is the asset that needs to be protected against side-channel attacks. RTL level functional simulation is performed to generate power profile/switching activity (toggle count) of design that are used as power traces. We also assume that there is a scheduler in place that controls the simultaneous operation of the additional IPs. For our measurement purposes, we use several benchmark circuits from ISCAS’89 suits which differ in terms of the number of inputs and outputs, gates, flipflops, etc.

\textbf{Workflow:} Figure \ref{fig:psc_methodology} illustrates the workflow of the PSC vulnerability measurement to provide a security score for an RT-level design of a system/subsystem. The measurement approach starts with the functional simulation of the crypto core and additional IPs. AES core is feed with a set of random plaintexts for , two keys from a set of predefined keys or random keys. The additional IP blocks perform random logic operations and add noise to the power trace. The keys are chosen such that each key consists of the same subkey, e.g, key0= 0x1515…15. For key1 and key2, hamming distance between two different subkeys is maximum.  For key1 and key2 with a number of random plaintexts, two sets of SAIF files are generated that contains the switching activities/toggle counts, considered as the power trace in the rtl level of abstraction. Toggle count for each key per input plaintext is stored into separate SAIF files (Switching Activity Interchange Format). Then the SAIF files are parsed through to generate switching activity distribution profiles for the subsystem and individual IPs for key1 and key2. Finally, JS divergence metric is calculated between the switching activity to assess PSC vulnerability of the subsystem at the RTL level.

\begin{figure*}[tbh]
 \vspace{0.5em}
    \centering
    \includegraphics[width=0.8\textwidth]{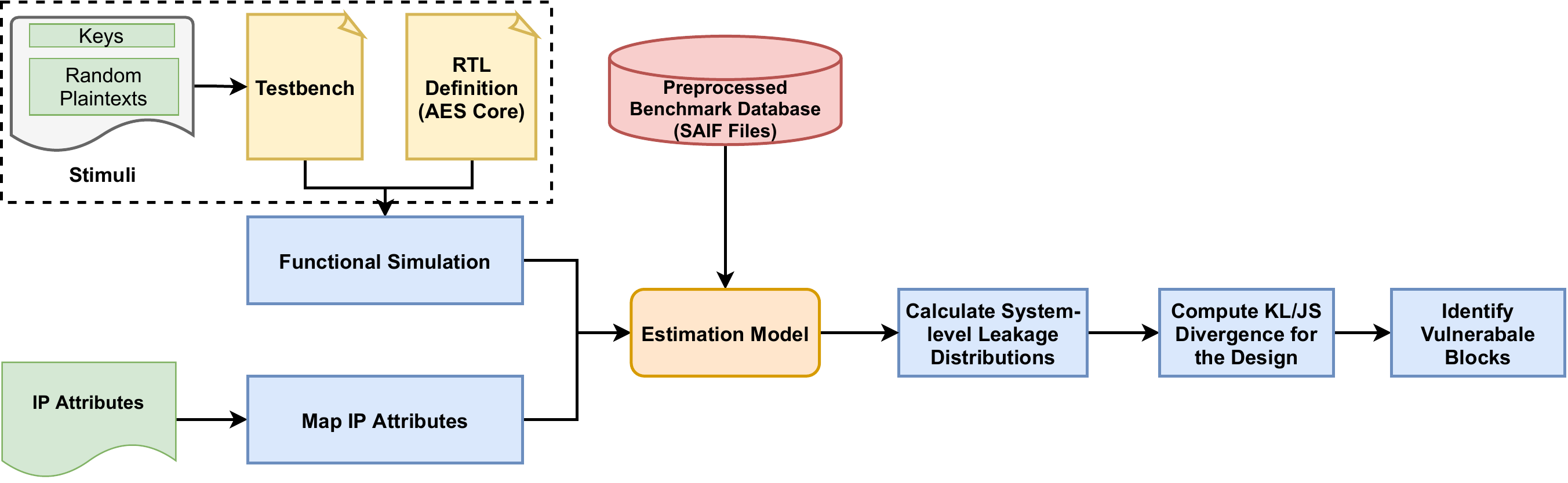}
 %   \vspace{-1em}
  %\vspace{-1.5em}
    \caption{PSC vulnerability estimation workflow.}
    \label{fig:psc_estimation}
   \vspace{-1.5em}
\end{figure*}

\textbf{Key-Pair Selection:} For the AES-LUT implementation, the encryption operation takes 11 cycle to finish. We collect switching activity traces for 1000 random plaintexts with two different keys for each clock cycle. The best key pair are among all possible key pairs is expected to provide the maximum KL divergence for vulnerability evaluation in the worst-case scenario. However, it is impossible and impractical to find the best key pair since the keyspace is huge, i.e., $\left(\begin{array}{c}2^{128}\\ 2\end{array}\right)$. So, the key pairs are selected empirically such that each key consists of the same subkey and hamming distance between two different subkeys is maximum. Table \ref{tab:psc_keypair} shows such a key pair used for AES crypto operations.

\begin{table}[b]
\vspace{-1.5em}
\caption{Sample key pair with maximum hamming distance.}
\label{tab:psc_keypair}
\centering
\begin{tabular}{|c|c|}
\hline
\textbf{$Key_1$} & $0$x$0000\_0000\_0000\_0000$                                      \\ \hline
\textbf{$Key_2$} & $0$x$FFFF\_FFFF\_FFFF\_FFFF$                                      \\ \hline
\end{tabular}
%\vspace{-2em}
\end{table}

\textbf{Evaluation Metric:} JS divergence metric estimates statistical distance between two different probability distributions. For instance, if power leakage probability distributions based on two different keys are distinguishable, JS divergence between these two distributions is high, which provides indication on how vulnerable the implementation is. The higher the difference between two power leakage distributions is, the higher impact the key has on the power consumption of the design/blocks, the more susceptible the design/blocks are to power analysis attack. For instance, if power leakage probability distributions based on two different keys are distinguishable, KL divergence between these two distributions is high, which indicates how vulnerable the implementation is. The JS divergence value for a design is translated to a security score between 1 to 5 based on predefined threshold value. The higher security score value refers to a more secure design.

\subsubsection{PSC Vulnerability Estimation}
Figure \ref{fig:psc_estimation} illustrates the workflow of the PSC vulnerability estimation to provide a security score for an RTL level design of a system/subsystem. Extensive simulation is not required for the estimation, rather each IP is mapped to an IP from a set of preprocessed IP database whose switching activity profiles are already generated. The mapping is performed based on the IP attributes such as number of inputs/outputs, d-flip flops, inverters and number of gates (AND, NAND, OR, NOR), etc. Then, the switching activity for the system is calculated from the AES core and mapped IPs. The rest follows the measurement process to calculate the JS divergence and provide an estimated security score.

\textbf{Pre-processed Benchmark Circuits:} Unlike measurement, the estimation  approach does not include the exhaustive simulation/emulation process. In order to map the IPs of the system to preprocessed IPs, we created a database with a number of IP from the ISCAS’89 benchmark suites. The properties of the benchmarks circuits are shown in Table \ref{tab:psc_benchmark}. The benchmark circuits are simulated with random input vectors to mimic the activity in a real system-on-chip, where the additional noises are uncorrelated to the crypto operations.

\begin{table*}[]
\centering
\caption{Pre-processed benchmark circuits used for sample subsystems.}
\label{tab:psc_benchmark}
\begin{tabular}{|c|c|c|c|c|c|c|c|c|c|}
\hline
\textbf{Bench Name} & \textbf{No. of Inputs} & \textbf{No. of Outputs} & \textbf{D-FF} & \textbf{Inverters} & \textbf{Gates} & \textbf{AND} & \textbf{NAND} & \textbf{OR} & \textbf{NOR} \\ \hline
s298                & 3                      & 6                       & 14            & 44                 & 75             & 31           & 9             & 16          & 19           \\ \hline
s344                & 9                      & 11                      & 15            & 59                 & 101            & 44           & 18            & 9           & 30           \\ \hline
s386                & 7                      & 7                       & 6             & 41                 & 118            & 83           & 0             & 35          & 0            \\ \hline
s400                & 3                      & 6                       & 21            & 58                 & 106            & 11           & 36            & 25          & 34           \\ \hline
s420                & 18                     & 1                       & 16            & 78                 & 14             & 49           & 19            & 18          & 34           \\ \hline
s444                & 3                      & 6                       & 21            & 62                 & 119            & 13           & 58            & 14          & 34           \\ \hline
s510                & 19                     & 7                       & 6             & 32                 & 179            & 34           & 61            & 29          & 55           \\ \hline
s526                & 3                      & 6                       & 21            & 52                 & 141            & 56           & 22            & 28          & 35           \\ \hline
s641                & 35                     & 24                      & 19            & 272                & 107            & 90           & 4             & 13          & 0            \\ \hline
s713                & 35                     & 23                      & 19            & 254                & 139            & 94           & 28            & 17          & 0            \\ \hline
s820                & 18                     & 19                      & 5             & 33                 & 256            & 76           & 54            & 60          & 66           \\ \hline
s832                & 18                     & 19                      & 5             & 25                 & 262            & 78           & 54            & 64          & 66           \\ \hline
s838                & 34                     & 1                       & 32            & 158                & 288            & 105          & 57            & 56          & 70           \\ \hline
s953                & 16                     & 23                      & 29            & 84                 & 311            & 49           & 114           & 36          & 112          \\ \hline
s1196               & 14                     & 14                      & 18            & 141                & 388            & 118          & 119           & 101         & 50           \\ \hline
s1238               & 14                     & 14                      & 18            & 80                 & 428            & 134          & 125           & 112         & 57           \\ \hline
s1423               & 17                     & 5                       & 74            & 167                & 490            & 197          & 64            & 137         & 92           \\ \hline
s1488               & 8                      & 19                      & 6             & 103                & 550            & 350          & 0             & 200         & 0            \\ \hline
s5378               & 35                     & 49                      & 179           & 1775               & 1004           & 0            & 0             & 239         & 765          \\ \hline
s9234               & 36                     & 39                      & 211           & 3570               & 2027           & 955          & 528           & 431         & 113          \\ \hline
s13207              & 62                     & 152                     & 638           & 5378               & 2573           & 1114         & 849           & 512         & 98           \\ \hline
s15850              & 77                     & 150                     & 534           & 6324               & 3448           & 1619         & 968           & 710         & 151          \\ \hline
s38417              & 28                     & 106                     & 1636          & 13470              & 8709           & 4154         & 2050          & 226         & 2279         \\ \hline
s38584              & 38                     & 304                     & 1426          & 7805               & 11448          & 5516         & 2126          & 2621        & 1185         \\ \hline

\end{tabular}
\vspace{-1.5em}
\end{table*}

\subsubsection{Results}

Now, in order to measure the system-level vulnerability of a design, we utilize the PSC vulnerability estimation framework to generate the switching activity of each IP with varying degrees of parallel activity. For a small subsystem with an AES core and few other IP blocks performing random logic operations, PSC vulnerability is affected by the additional noises in the power trace introduced by the platform level parameters that translate to additional toggle count in the RTL level. In our preliminary measurement of a small subsystem, we present the effects of these additional switching activities.

\textbf{Measurement Results:}
With our framework, we initially perform the functional simulation of a system with the AES-LUT and 4 benchmark circuits with the stimulus as mentioned above. From the generated SAIF files, the distribution of switching activity is generated for individual IPs, shown in Table \ref{table:switching_activity}. Here each of the rows corresponsd to the toggle count during a encryption operation on a random plaintext for a given key. Column A shows the samples of toggles count for the subsystem and column B to F shows the toggle counts of AES core and 4 benchmarks. The additional switching activities increases the noise in the distribution. The switching activity (toggle rate) of the sample subsystem and individual IPs are calculated for each clock cycle and each plaintext with $key_1$ and $key_2$. The two sets of data are used to compute the Kullback-Leilbler or Jensen-Shannon divergence to find similarity between the two distributions.

\begin{table}[b!]
\vspace{-1em}
\caption{Sample of switching activity profile of a subsystem and individual IPs for a given key.}\label{table:switching_activity}
\begin{tabular}{|c|c|c|c|c|c|}

\hline
\textbf{Subsystem   (uut)} & \textbf{AES core} & \textbf{s832} & \textbf{s953} & \textbf{s1488} & \textbf{s5378} \\ \hline
22379                      & 11580             & 242           & 1314          & 291            & 8213           \\ \hline
22532                      & 11622             & 250           & 1317          & 311            & 8282           \\ \hline
22750                      & 11496             & 323           & 1313          & 370            & 8498           \\ \hline
22749                      & 11586             & 325           & 1315          & 294            & 8483           \\ \hline
22397                      & 11388             & 255           & 1311          & 294            & 8464           \\ \hline
22555                      & 11461             & 261           & 1313          & 367            & 8421           \\ \hline
22752                      & 11614             & 243           & 1311          & 369            & 8459           \\ \hline
22579                      & 11569             & 249           & 1311          & 369            & 8459           \\ \hline
22528                      & 11470             & 260           & 1313          & 313            & 8395           \\ \hline
22512                      & 11428             & 253           & 1311          & 309            & 8497           \\ \hline
\end{tabular}
%\vspace{-1.2em}
\end{table}

\begin{figure}[]
 %\vspace{-1.5em}
    \centering
    \includegraphics[width=\columnwidth]{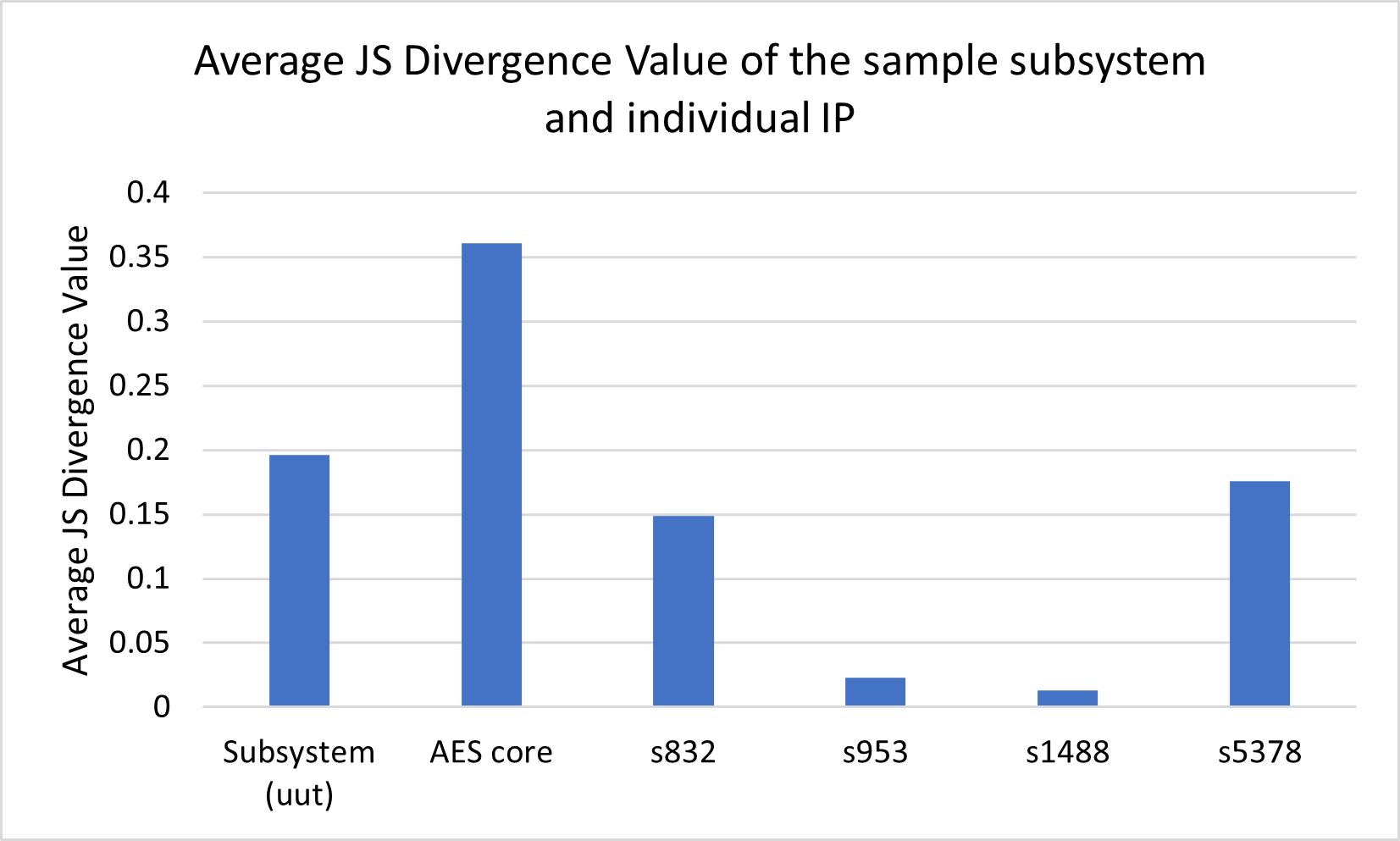}
 %   \vspace{-1em}
  %\vspace{-1.5em}
    \caption{PSC vulnerability measurement results for a sample subsystem.}
    \label{fig:result_psc_measurement}
   \vspace{-1.5em}
\end{figure}

The calculated JS divergence matrix for each clock cycles from key1 and key2 is shown in Figure \ref{fig:result_psc_measurement}. From the KL divergence matrix, we observe that the AES module has relatively higher correlation of power traces with keys compared to benchmark IPs. It is expected as the benchmark IPs are based on random inputs, they show very low level of JS divergence value. Now, if we look at the JS divergence values of the sample subsystem, we can see that the additional noises reduce the correlation significantly. %The red blocks are the most vulnerable to power side-channel attacks, the reddish-yellow blocks have moderate level of vulnerability, and the green blocks have low level of vulnerability.

\textbf{Estimation Results: }For estimation, individual AES implementations are preprocessed to generate the switching profiles. Next, we use the IP mapping with the input metadata of the IPs and form the composite distribution of switching activities based on the additional noise level. Here, in Table \ref{tab:psc_estimation}, the average JS divergence  value of 3 different AES implementation is presented with different number of additional IPs.
For the demonstration purpose, we estimate the power trace for a few sample subsystems with different degrees of additional noises. As we move to the right, more random activities are added to the AES switching activities which decreases the JS divergence value as it successfully hides the correlation of power traces.

% \usepackage{multirow}
\begin{table}[b]
\vspace{-1em}
\caption{Estimated JS divergence matrix for multiple AES implementations with varying number of IPs.}\label{tab:psc_estimation}
\begin{tabular}{|c|c|c|c|c|}
\hline
\multirow{2}{*}{\textbf{AES Cores}} & \multicolumn{4}{c|}{\textbf{Estimated JS Divergence Value for subsystems}} \\ \cline{2-5} 
                                    & \textbf{AES}   & \textbf{AES+2IPs}   & \textbf{AES+4IPs}   & \textbf{AES+6IPs}  \\ \hline
AES-LUT                             & 0.3125              & 0.2353              & 0.2012              & 0.1174             \\ \hline
AES-GF                              & 0.2962              & 0.2877              & 0.2700              & 0.2457             \\ \hline
AES-CTR                             & 0.1881              & 0.1897              & 0.1861              & 0.1302             \\ \hline
\end{tabular}
%\vspace{-1.5em}
\end{table}

\textbf{Limitations and Scopes:}
In this demonstration, we are presenting the proof of concept for PSC score measurement and estimation for small subsystems. We demonstrate how additional noises from IP-level parallel activities impact the vulnerability score in a system. However, the estimation model provides very crude approximation of noise as the input attributes of IPs are limited. With additional information such as the IPs that share the power rail, the timing overlap of the IPs, etc, a robust estimation can be modeled to estimate the additional noise level from the metadata of the IPs provided as user input. Then, the composite switching activity profiles for a platform can be generated by adding appropriate number of benchmark IPs and tuning the weights based on the input platform level parameter. In order to validate the estimation score, extensive simulation measurements of the platform will be required.

\section{Security Optimization}
There have been a lot of research for power, area, and performance optimization of a design \cite{chuang1995timing, schliebusch2005optimization, ahmed2015modified, saha2015survey, tehranipoor2005low}. However, the security optimization has been explored till date. This section describes a new concept called \textit{Security Optimization}. Power, Performance, and Area (PPA) optimization has already been an essential part of the design flow. Existing industry-level EDA (Electronic design automation) tools contain the feature to optimize these three attributes. However, there is no such concept of security optimization to date, and thus, the tools are not equipped with the security optimization methodology. The fundamental difference between security and the other three attributes: power, performance, and area, is that these three attributes are one-dimensional while security is multidimensional. For example, in the perspective of this paper, the term security is composed of five different threats: IP Piracy, Power Side-Channel Analysis, Fault-Injection, Malicious Hardware, and Supply-Chain. Researchers have always been considering building countermeasures, detecting and fixing vulnerabilities for individual threats. To date, there is no comprehensive solution that focuses on improving security as a whole, given the list of threats to be considered. 

We argue that the security against an individual threat is not entirely orthogonal to others. Security improvement against one threat may weaken the security against others, thus making security optimization a non-trivial task. For example, let's consider the fact described in section {\ref{sec:reEstimation}}, different DFT features (i.e decompressor/compressor) greatly benefit security against IP piracy. However, on the other hand, DFT features might facilitate malicious hardware attacks by leaking secret information. Thus, it is important to develop a methodology to get optimum combined security against the threat model. 

Towards this approach, the first and most important task is to identify which parameters' contribution to the security expands across the threat space. For example, DFT is a parameter that affects security against both IP piracy and Malicious hardware attack. The second step is to identify the impact factor of the parameters on security enhancement or degradation against individual threats. As shown in section {\ref{sec:reEstimation}}, the DFT features decompressor and compressor enhances the robustness of security against IP piracy by a specific amount. Similarly, the quantification of security degradation against malicious hardware by information leakage through DFT should also be calculated. Thus, to get optimum security, threats should not be considered individually; instead, a collective approach should be taken to achieve security against all threats without benefiting or hampering the security against any particular threat.  

\begin{comment}
\textcolor{blue}{
Discuss the necessity of security optimization. 
Focus that this concept is not yet been explored. 
Researchers only focus on individual threat; however, security is not a single term like power, area or performance used in design optimization.  
Discuss with example that improving the security for one threat may weaken the other threat.  }
\end{comment}

%\input{challenges}
\section{Challenges}

This section presents several challenges during estimation, measurement, and optimization of security at the platform level. We list out the challenges in three categories:   1) Challenges in Platform Level Security Estimation and Measurement, 2) Challenges in Achieving Accurate Estimation, and 3) Challenges in Security Optimization.

\subsection{Challenges in Platform Level Security Estimation and Measurement}
The major challenges in estimating and measuring the platform level security can be itemized as follows: 

\begin{enumerate} 
\item Identifying the representative parameters contributing to the platform level security.
\item Modeling the impact of the platform level parameters on platform level security estimation.
\item Introducing the platform-level parameters with the design while performing the actual attack during measurement.
\end{enumerate}

During the transition from IPs to platform, several new parameters are introduced at different stages and abstraction levels of the design flow. However, not all these parameters impact the system's security. Identifying the parameters that impact security against different threats is a challenging task, as there is no systematic approach to accomplish it. Modeling the impact of different platform-level parameters on security is also a non-trivial task. A list of challenges for different threats is listed below.

\noindent
\textbf{IP Piracy: }
\begin{itemize}
    \item The RT level platform does not provide information about the test structure introduced in later design stages. The robustness of a platform against IP piracy through SAT attack greatly depends on the test structure of that platform. Hence, estimating the security based on parameters that do not exist becomes quite challenging.
    
    \item As discussed in Section \ref{sec:caseStudy}, the proposed estimation model for IP piracy is built based on a data-driven model, in which a set of data for different designs and the values of the platform-level parameter (in this case, compression ratio) are used to predict/estimate the security of an unknown design. However, selecting a suitable set of such designs and the values of the platform level parameters is difficult.  
    
    \item While estimating an unknown platform's security against IP piracy, the first step is to match the metadata of that design with all the designs used in modeling. A bad choice of metadata can potentially steer the estimation method toward a poor security estimation outcome. However, developing a representative set of metadata is a very challenging task. Also, with the developed metadata, the simple cosine similarity-based matching algorithm does not accurately select the suitable model, which leads to the need for future machine learning (ML) based approaches.
    
\end{itemize}

\noindent
\textbf{PSC Analysis: }%\todo[inline]{To be completed by Kawser}

\begin{itemize}
    \item Parallel activities on the system bus and active IPs that share the power rail with the target IP are the most critical elements in platform-level PSC vulnerability assessment. Without extensive human effort from verification engineers, identifying and controlling IP activities in complex SoC designs becomes extremely challenging. Moreover, HW/SW co-verification for a platform will be required to compute the simulated power traces required for PSC vulnerability estimation and measurement. 
    
    \item Because RT level designs do not include any physical information about the blocks, the simulated power model may fail to capture factors such as shared power distribution network, voltage regulator, clock jitters, etc. Analysis at a lower abstraction level (e.g., gate-level, layout-level) may improve the accuracy of the power models. However, each abstraction level increases the analysis time by tenfold, making it nearly impossible to perform gate-level, or layout-level analysis for a full-blown SoC design.
    %\item It is challenging to include platform-level factors such as power   improve accuracy  making it challenging to compute accurate power traces.
    \item The data-driven model proposed for fast estimation of PSC vulnerability may not be able to predict accurate switching activity of the IPs. It will be significantly reliant on the IP properties chosen to match the switching activities of the existing IP repository. ML-based algorithms could be useful for mapping metadata in order to estimate switching activity of the IPs. However, the training phase will have to incorporate power models for a large number of IPs with varying workloads, temporal overlapping, etc. This becomes a challenging task due to the sheer volume of the training data and time required to train the model. 
    
    \end{itemize}

\noindent
\textbf{Fault Injection: } 

\begin{itemize}
    \item Modeling of physical parameters involved with fault injection methods, such as laser, clock, voltage, etc., is the most challenging part when it comes to assessing fault injection susceptibility of a pre-silicon design/SoC.
    \item Considering the large size of SoCs, the possible number of fault locations grow exponentially in the design/SoC. Therefore, exhaustive simulation of all fault nodes and analyzing their impact on the design's security is a challenging part.
\end{itemize}

\noindent
\textbf{Malicious Hardware: }
\begin{itemize}
    \item A major platform-level parameter contributing to the platform-level security against malicious hardware is the set of security policies. It is very unlikely that an exhaustive set of security policies will be available at the earlier design stages, which significantly hinders the estimation capability. Even with a set of security policies, quantifying the contribution to platform-level malicious operation is a daunting task.
    
    \item The interaction probability between IPs significantly impacts the security of the entire platform. An IP with the security asset having a high interaction probability with a malicious IP can lead to a greater risk of malicious operation than the interaction with a non-malicious IP. However, getting information about this parameter at an earlier platform stage is a challenging task.  
    
\end{itemize} 

\noindent
\textbf{Supply Chain}

\begin{itemize}
    \item As discussed in the previous section, the widely used counterfeit IC detection mechanism utilizes the security primitive PUF. As PUF is implemented using the process variation of the chip, it is practically impossible to measure the security against cloning attacks at the RT level, as there is no notion of process variation in this stage of the design.
    
\end{itemize}

\subsection{Challenges in Achieving Accurate Estimation}
The security estimation should provide a highly accurate estimate of the platform-level security, closely resembling the security measured in a full-blown SoC in silicon. The challenge is that the estimation is performed at the earlier stages (e.g., RTL, gate-level), and we aim to estimate the security at the SoC in silicon. Consider estimating the platform-level security against PSC attacks at the RT level. It is quite challenging to accurately model the impact of the platform-level parameters, such as PDN, Decap, DVFS, on the security at the RT level. These parameters are not available at the RT level and become available later in the design flow. The only information about the power traces that can be achieved at this stage is the switching activity, which potentially leads to a poor estimation. Thus, estimating the security at RT level closely representative of the platform-level security at silicon is a very challenging task. This is mainly because most platform-level information is unavailable at the RTL or gate level. 

\subsection{Challenges in Security Optimization}
Security optimization is another challenge as improving security against one threat may impact the robustness against the other threat models. Unlike power, performance, and area, security is not a single term; instead, it comprises a set of diverse threat models. For example, if a particular chip is claimed to be x\% secure, it does not provide specific information. Does it mean that the chip is x\% secure against all possible attacks? A chip can not be claimed secure against all threats. It is because the threats are not confined to a particular number. There are already a large number of threats available, and many more threats are yet to come. Thus, as a first step, the threat models should be clearly defined before estimating and measuring the security of a design. The security against power side-channel attacks may be orthogonal to the security against other threats such as IP protection through logic locking. Hence, the techniques to measure security against different threats should be developed separately, but the impact of one on another threat model must be measured or estimated within the SoC as well.

\section{Conclusion}
This work proposes an approach to develop the SoC-level security measurement and estimation. We discuss how the transition from  IP to SoC affects the overall SoC security. We identify additional parameters introduced during the SoC integration and their possible impact on developing the SoC-level security metric. We also present the step by step procedure for the measurement and estimation of the security against two threats: IP Piracy, and Power Side-Channel Analysis. We also discuss the major challenges that need to be addressed to obtain the full benefit of this approach. This work opens up new research directions that require more attention from the hardware security community.

\bibliographystyle{IEEEtran}
\bibliography{IEEEabrv,mybib.bib,lfi.bib}

\end{document}